Work dedicated to the centenary of the birth of L.S.Palatnik-scientist,teacher,citizen.

# STUCTURE AND QUALITY OF MONO- AND BI-COMPONENT FILMS,FORMED FROM ION-PLASMA STREAMS, TRANSFORMATED DURING THE INTERACTION WITH THE CONDACTIVE TARGET.

## A.S. VUS

National technical University «Kharkov polytechnic institute»
UKRAINE

*The essence of the method of transformation of ion-plasma streams was disclosed. A phenomenological model of binary alloy films formation was suggested, in the case of a conducting target being sputtered by ions of nongaseous substances. By example of Nb films formation and Nb-Sn system, the effect of kinetic and thermodynamic parameters on the formation processes of mono- and bi-component films, in the framework of the used method, were studied. The range of energies ($E_0 < E \leq 2E_d$, $\langle E \rangle \sim E_d$) of condensing particles, within which form highly stable mono-component films and coatings, was specified. The mechanisms of condensation-induced and radiation-induced diffusion were compared, indicating the energy ranges, which determine a predominant implementation of the vacancy or interstitial mechanism of condensation-induced diffusion.*

## 1. INTRODUCTION

From the moment of publication of earliest research papers, dedicated to the practical formation of fine films of pure metals and alloys during the sputtering of metal targets by accelerated ions of metals [1,2] , had past more than 20 years, however there are extremely not enough published works dedicated to this theme in an open press. Among foreign researchers our proposed method has been implemented by the authors of work [3] in the form of obtaining the mono-component films. In recent years can be found isolated occasional works of other authors, dedicated to the phenomenon of self-sputtering and reflection of ion-plasma streams by using the vacuum-arc film and coatings deposition method ,see example, [4-6].

One of the objectives of this paper is to address the lack of information about the proposed method of transformation of ion-plasma streams and features of the formation in vacuum of mono- and bi-component films. Transformation of initial ion-plasma stream take place as a result of its interaction with a conductive target, which has certain electrical potential and can be located in the magnetic field of a special configuration. The method allows to conduct the formation of films and coatings in super-high vacuum conditions.

In the previously described device [1], during the sputtering of a conductive target with accelerated metal's ions, transport of ions from the cathode of vacuum arc device to a target at a distance of ~0,5 m is carried out by means of plasma guide. This significantly reduces the coefficient of efficiency of the device.

Cathodes, made of refractory metals, generate plasma streams with a high degree of ionization and low content of condensed (drop) phase (at the level of several percent). Therefore, in obtaining an alloy with a high concentration of component, supplied by the cathode, the target can be positioned opposite the cathode. It is obvious that the performance of the spray device can be significantly increased.

Part of this work, dedicated to the formation of mono-component films, is largely a compilation of the results, obtained earlier by the author. Study of the formation of two-component films was carried out on the example of niobium-tin system, with the location of a tin target within the line of sight to the niobium cathode. Selection of the Nb-Sn system is due to the presence of the superconducting compound $Nb_3Sn$, as well as the possibility of studying the formation of the alloy, the components of which vary greatly in their melting point and saturated vapor pressure. Management of the structural state, phase composition, and ultimately the properties of the mono- and bi-component films was carried out by specifying the required ranges of kinetic and thermodynamic parameters of the ion-plasma streams transformation system.

Study of the formation of films of niobium-tin system was started more than 20 years ago, but there was no publication on this topic, as initiated research was soon interrupted in conjunction with the more relevant at that moment research work in the field of high temperature superconductivity. One of the developed by us techniques in that period solved the problem of formation of stoichiometric composition of metal subsystem HTS (high temperature superconductivity) films during the sputtering of multi-component target by accelerated multi-component ion stream [7,8]. The focus of previous studies had been directed at obtaining the actual material, that was used to adjust the parameters of the sputtering system, while insufficient attention was paid to the creation of a consistent sputtering model of a conductive target by the ions of metals.

## 2. THE MODEL OF THE ALLOY FORMATION

Time-constant flow of substance (A) falls on the surface of the target (B) during the stationary arc discharge in vapors of the cathode material (A) (Fig.1). Flux density of substance A ($J_A$) is the sum of the three components of the stream:

$$J_A = J_{A,i} + J_{A,0} + J_{A,\kappa}, \tag{1}$$

where i- ions, o-neutral atoms, κ- micro drops, electrons do not fall on the target at its negative potential.

Accelerated ions bombard the target and spray a layer of implantation Δ, formed by type A atoms. A layer of implantation Δ with a reasonable degree of accuracy can be considered equal to the average projective run of the bombarding ions, which have maximum charge (max $\langle R_P \rangle$). There is an accumulation of element A in the layer of implantation, which occurs during the sputtering process at a constant accelerating voltage on the target (U=const) and constant over time flux density of cathode substance ($J_A$= const). The concentration profile of the element A tends to its stationary state.

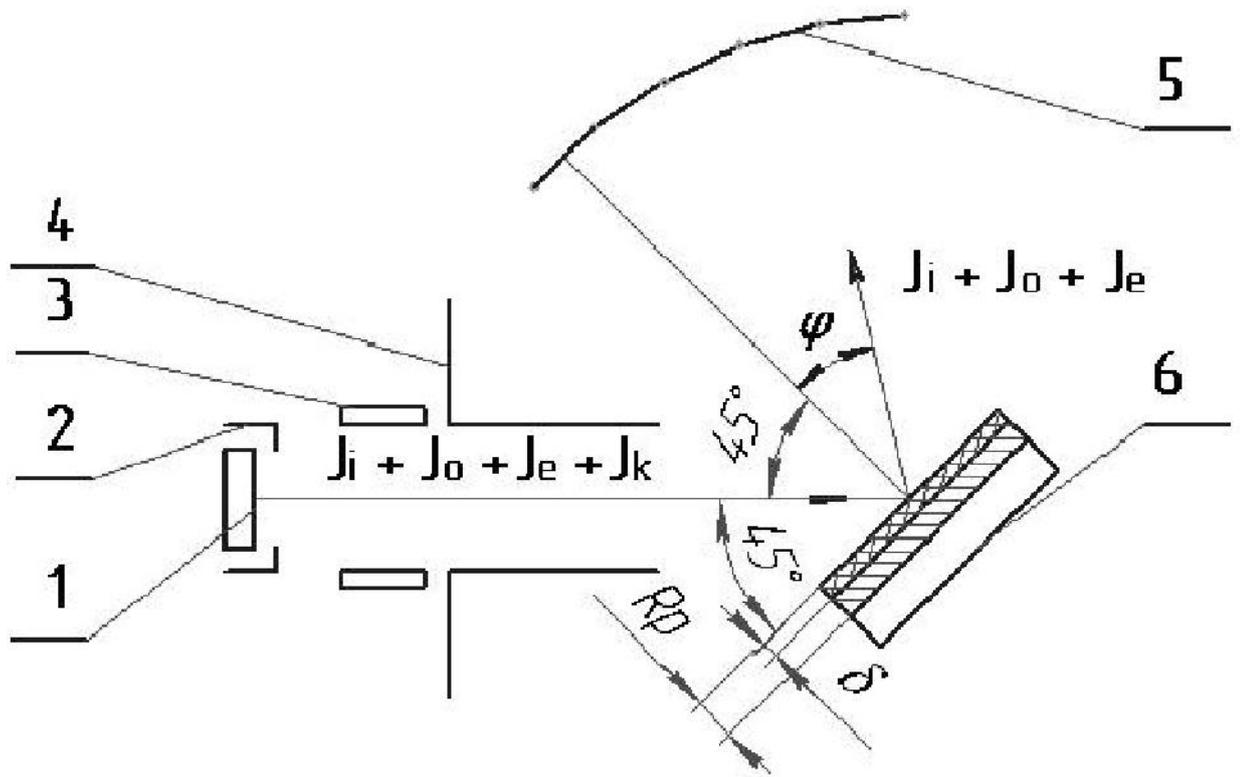

Fig.1. The experimental scheme. The components of initial and transformed flow. J – ions, 0 – neutral atoms, k – microdrops, e – electrons. 1- cathode; 2- electrostatic screen, 3- anode, 4- diaphragm, 5 – substrates, 6- target.

Calculation of the concentration of implanted ions, taking into account the movement of the sputtered target surface under the assumption that the integral sputtering coefficient of material (S) and the average projective range do not depend on the composition, is given by the expression [9]:

$$c(x,t) = \frac{1}{S} \int_{x}^{x+J_{A,i}St/n_0} g(z)dz, \qquad (2)$$

where $n_0$- density of the target atoms, t- the total duration of implantation, g(z)- normalized per unit area implantation profile without taking into account the sputtering.

From (2) implies that on the surface of the target the concentration of the implanted material tends to $c_F \rightarrow 1/S$ when $t \rightarrow t$.

The presence of multiply charged ions and neutral componenta in the stream $J_A$ contributes to the formation of almost П-shaped stationary concentration profile of implanted impurity. This is due to the superposition of concentration profiles of the incorporated into the target recoil atoms and ions, whose charge varies from $e$ to $ne$, for niobium n = 5 [10]. The use of implantation of multiply charged ions is a special case of poly-energy ion implantation method, which is used to create a homogeneous distribution of impurity to a certain depth followed by a sharp decline in concentration to zero [11].

Sputtered streams of atoms A and B ($J_{A,S}$ and $J_{B,S}$), generated by a layer of output Δ can be used to obtain films of the alloy A-B constant or variable concentration, of the vapor elasticity of components, forming the alloy. Presence of a stationary concentration profile of element A in the implanted layer causes the consistency in time of streams $J_{A,S}$ and $J_{B,S}$.

In the absence of droplet phase, the main problem in obtaining an alloy with regulated concentration is an inertia of sputtering system, which is mainly determined by the time out of the sputtering process on the stationary regime, when the accelerating potential on the target is changing.

Introducing the outcome probability of the atoms A and B from a sputtering layer $W_A$ and $W_B$ respectively, for the flux density of the sputtered atoms can be written:

$$J_{A,S} = J_{A,i}\, S_A\, W_A = J_{A,i}\, S_{A,P} \qquad (3)$$

$$J_{B,S} = J_{A,i}\, S_B\, W_B = J_{A,i}\, S_{B,P} \qquad (4)$$

$S_{A,P}$ and $S_{B,P}$ – partial coefficients of sputtering of components A and B.

We assume that the accelerating potential of the target is large enough and that the bombarding ion necessarily lead to the dispersal of the substance, ie

$$W_A + W_B = 1 \qquad (5)$$

The probability of outcome of the sputtered atoms is directly proportional to their concentration ($c$) in the layer of outcome δ, with the coefficients of proportionality equal to one by the normalization condition ($c_{A,\delta} = W_A$, $c_{B,\delta} = W_B$).

Integral coefficient of sputtering is determined as follows:

$$S = \frac{J_{A,S} + J_{B,S}}{J_{A,i}} = S_A W_A + S_B W_B = S_{A,P} + S_{B,P} \qquad (6)$$

In the process of entering the stationary sputtering regime coefficient of sputtering may vary not only when the concentration of components changes, but also with the possible formation of compound of elements A and B in the layer of implantation, as in this case, the atomic bond force in the lattice varies.

Duration of output at a stationary sputtering regime depends on the transport processes (diffusion, segregation), preferential sputtering of components, development of the A layer of mono-component target, involves the use of several simplifying assumptions:

1. Believe the target is isotropic, with an infinite thickness.
2. Not consider the possibility of formation of compounds of elements A and B, the segregation and nucleation of new phases in the layer of implantation.
3. Ignore the speed of diffusion "degradation" of the layer of implantation compared with the rate of displacement of the target surface.
4. Not taken into account the development of the surface relief and preferential sputtering of the components.
5. Micro drops of the substance A on the target surface is a mono-component part of a target with some effective thickness.
6. Dilatational effects in the atomic implantation layer of the element A in the target are absent.

## 2.1 ASSESSMENT OF THE IMPACT OF A DROPLET PHASE ON THE PROCESS OF SPUTTERING OF THE TARGET.

### 2.1.1 SPUTTERING TIME OF AN INDIVIDUAL DROP.

For simplicity, we assume that the micro droplets are disc-shaped. Dispersion of a micro droplet, situated on the surface of the target is a process of self-sputtering of the compound A. Ignoring the process of reflection of atoms A, and taking into account the processes of sputtering and condensation, for a drop of unit area we can write the balance of particles at its full dispersion:

$$(J_{A,S} - J_A)\tau_V = \frac{\rho_A h}{\mu_A} N_0 \qquad (7)$$

In which $\tau_V$ -the lifetime of the drop, $\rho_A$ -the density of substance A, $\mu_A$ -the molar mass of a substance A, $h$ -the thickness of the drop, $N_0$ -Avogadro's number.

Introducing the coefficient of ionization of atoms A stream, generated by the cathode of the vacuum arc $\gamma = J_{A,i} / J_A$, from (7), taking into account (6), we obtain:

$$\tau_V = \frac{\gamma h \rho_A N_0}{\mu_A J_{A,i}(\gamma S_{A,0} - 1)}, \qquad (8)$$

Where $S_{A,0}$ –the coefficient of self-sputtering of the substance A.

When U= 1 kV; $\langle q \rangle \approx 3|e|$ [10] ($\langle E \rangle$ = 3000 eV) ; $h \sim 10^{-4}$ cm; $\gamma$ = 0,9; $S_{A,0} \sim 2,2$ [12] ($S_{A,0}$ estimated from the data of sputtering of niobium by ions of krypton), $J_i$ = 2.5 $10^{16}$ cm$^{-2}$ sec$^{-1}$ we obtain, that $\tau_V \sim 200$ sec.

### 2.1.2 TIME DEPENDENCE OF THE ESTABLISHMENT OF STEADY-STATE DISTRIBUTION OF DROPLET PHASE.

Introducing the coefficient of filling per unit surface area by drops (*k*), the growth rate of the coefficient of filling η = const, which is determined by the parameters of the arc evaporation system and, taking into consideration that, 1/τ$_v$ is the sputtering rate per unit surface area of a niobium micro drop, the change of the coefficient of filling over time (*dt*) can be written as:

$$dk = (\eta - \frac{k}{\tau_V})dt \qquad (9)$$

With the boundary conditions:

$$\begin{cases} t = 0, \ldots k = 0 \\ t \to \infty, \ldots k \to K \end{cases}$$

Where *K* – stationary value of the coefficient of filling of the target by drops. Solving (9), we find:

$$k = \tau_V \eta \left[1 - - \exp\left(-\frac{t}{\tau_V}\right)\right] \qquad (10)$$

Hence $K = \tau_V \eta$.

Carried out using РЭМ – 200 study of the surface morphology of niobium films, condensated during different time intervals on aluminum substrates, fixed on the target surface, at its zero potential gives the growth rate of the coefficient of filling of the target by niobium micro drops $\eta \sim 2{,}5 \cdot 10^{-3}$ sec$^{-1}$, so that the stationary value of the coefficient of filling of the target is $K \sim 0{,}5$ when $U = 1$ kV.

From (10) follows that the surface coverage of the target by micro droplets, which differ from the stationary distribution by 0,1 %, while the lifetime of the drop 200 sec, achieved in about $1{,}4 \cdot 10^3$ sec.

## 2.2. MODEL OF OBTAINING AN ALLOY WITH A CONSTANT CONCENTRATION OF THE COMPONENTS ALONG THE FILM THICKNESS.

We estimate the time- to- mode dispersion ($t_x$), in which the concentration of element A in the surface layer is different from the stationary value on a given (arbitrarily small) amount ($x$), when accelerating potential on the target spasmodically changes, eg when it is switched on.

While accept simplifying proposals of the above.

Time ($t_x$) can be estimated from (2), using the specific type of implantation profile, the approximate time dependence of the yield on the regime can be determined by solving a system of differential equations, describing the balance of particles of type A and B in the implantation layer of unit area. And a reference system is convenient to associate with a moving target surface:

$$\begin{cases} \dfrac{dN_A}{dt} = J_A - J_{A,S} - J_{A,R} \\ \dfrac{dN_B}{dt} = J_{B,Q} - J_{B,S} \end{cases} \quad (11)$$

Where $\dfrac{dN_A}{dt}$ and $\dfrac{dN_B}{dt}$ the rate of change in the number of particles of type A and B in the implantation layer of unit area, $J_{A,R}$ - the flux density of the reflected from the target surface particles of type A, $J_{B,Q}$ - density of the "quasi-stream" of the matrix atoms in the implanted layer, determined by the speed of the inner boundary movement of the modified layer $v(\Delta)$.

We are interested in small deviations from the stationary state of the sputtering process.

Functions describing the instantaneous values of specific variables in the layer of implantation, and the functions of the same name, averaged over the volume of this layer, have a common time asymptotes. Taking into consideration (3) and (4), from (11) we obtain:

$$\begin{cases} \left\langle \dfrac{dN_A}{dt} \right\rangle = J_A - J_{A,R} - J_{A,i} S_A \langle W_A \rangle \\ \left\langle \dfrac{dN_B}{dt} \right\rangle = \langle n_{B,V} \rangle v(\Delta) - J_{A,i} S_B \langle W_B \rangle \end{cases} \quad (12)$$

We express $\langle W_A \rangle$ through a number of particles of type A in the implementation layer and the total number of particles ($N$) in this layer:

$$\langle W_A \rangle \approx \frac{\langle N_A \rangle}{\langle N \rangle} \tag{13}$$

Assuming that $\langle N \rangle \approx$ const, we obtain instead of system (12) independent differential equations. Leaving the signs of averaging, we can write:

$$\frac{dN_A}{dt} = \frac{1-\chi}{\gamma} J_{A,i} - \frac{N_A}{N} S_A J_{A,i} \tag{14}$$

Where $\chi = J_{A,R}/J_A$ - the reflection coefficient of the incident flux of atoms on the target. Solving (14) with the boundary conditions:

$$\begin{cases} t = 0, \ldots N_A = 0 \\ t \to \infty, \ldots N_A \to N_A^* \end{cases} \tag{15}$$

We find:

$$N_A = \frac{1-\chi}{\gamma S_A} N \left[ 1 - \exp\left(-\frac{S_A J_i}{N} t\right) \right] \tag{16}$$

Stationary dose of implanted atoms of type A per unit area of the target defined by the expression:

$$N_A^* = \frac{1-\chi}{\gamma S_A} N \tag{17}$$

We estimate the time-to-mode dispersion, in which x = $N_A / N_A^*$ = 0.999, for the following experimental data on the sputtering of a tin target with accelerated ions of niobium: $J_i$ = 2.5 $10^{16}$ cm$^{-2}$ sec$^{-1}$, $U$ = 1000V, $\langle q \rangle \approx 3|e|$, $R_p$ = 2.8 nm, [13], $S_{Nb} \sim 2,2$, $\mu_{Sn}$ = 119 g/mol, $\rho_{Sn}$ = 7,3 g/cm$^3$. From (16) without taking into account the reflection of the ion flux, we obtain that $t(x) \sim 1,3$ sec

Under steady-state sputtering from (12) we obtain:

$$\begin{cases} 0 = J_A - J_{A,R} - J_{A,i} S_A c_{A,\delta} \\ 0 = \langle n_{B,\Delta} \rangle v(\Delta) - J_{A,i} S_B c_{B,\delta} \end{cases} \tag{18}$$

Where $c_{A,\delta}$ and $c_{B,\delta}$ concentration of the components A and B on the target surface (concentration of components in the output layer δ), and $v(\Delta) = v(0)$, i.e. upper and lower boundaries of the implantation layer move with equal speed.

From (18) we obtain that the partial sputtering coefficient of the compound A is equal to:

$$S_{A,P} = \frac{1-\chi}{\gamma} \qquad (19)$$

Flux density generated by the target from the field free of the drop phase, determined as $J_T = \langle n \rangle v(0)$, or $J_T = J_{A,i} S$ Then from (18), taking into account that $\langle n_{B,\Delta} \rangle = c_{B,\Delta} \langle n \rangle$, we obtain:

$$S = S_B \frac{c_{B,\delta}}{c_{B,\Delta}} \qquad (20)$$

Since $c_{A,\delta} = W_A$, and $c_{B,\delta} = W_B$, then from (6) and (20) it follows that:

$$S = S_A \frac{c_{A,\delta}}{c_{A,\Delta}} \qquad (21)$$

Therefore

$$\frac{S_A}{S_B} = \frac{\dfrac{c_{A,\Delta}}{c_{B,\Delta}}}{\dfrac{c_{A,\delta}}{c_{B,\delta}}} \qquad (22)$$

Sputtering produced by ions of substance A, therefore $S_A \equiv S_{A,0}$. Taking into account (21), from (18) we obtain:

$$S = \frac{1}{c_{A,\Delta}} \frac{1-\chi}{\gamma} \qquad (23)$$

Let us find the integral sputtering coefficient from the area free of micro droplets in the presence of the target surface of a droplet phase. Since $c_{B,f} = \dfrac{(1-K)J_i S_B c_{B,\delta}}{(1-K)J_i S + K J_i S_{A,0}}$, and taking into account (20) and (21), for the integral sputtering coefficient we obtain:

$$S = \frac{1-\chi}{\gamma} \frac{1}{c_{A,f} - c_{B,f} \dfrac{c_{A,\Delta}}{c_{A,\delta}} \dfrac{K}{1-K}} \qquad (24)$$

We express the integral sputtering coefficient through the current on the cylindrical target ($I_T$) of radius R and film growth rate $v_f$, in the presence on the target surface of droplet phase. The flux of the substance, sputtered from elementary area of the target $rdrd\varphi$, at a distance $r$ from its center, condenses on the site of the substrate $\delta F$, which is located on the axis of the target at a distance $L$ from its surface. Element of the film surface sets a solid angle $d\varpi = \delta F / (L^2 + r^2)$.

We assume that the sputtered particles have a cosine distribution of the emission direction. Easy to show that the maximum density of the sputtered flux $J_{max}$ is associated with an average density of the sputtered flux $\langle J \rangle$ as follows: $J_{max} = \pi/2 \langle J \rangle$ and $\langle J \rangle = S J_i$.

Neglecting the reflection of sputtered particles from the film surface, the balance of sputtered on the substrate particles, we can write as:

$$\frac{\pi}{2} J_i [KS_{A,0} + (1-K)S] \int_0^R \frac{L^2 r dr}{(L^2 + r^2)^2} \frac{\delta F}{2\pi} \int_0^{2\pi} d\varphi \int_0^\tau dt = \frac{\langle \rho_f \rangle \delta F h N_0}{\langle \mu_f \rangle} \quad (25)$$

where $\langle \rho_f \rangle$ and $\langle \mu_f \rangle$ the average density and average molar mass of the film substance.

From (25) taking into account the fact that $J_i = \dfrac{I_T(1-\beta)}{\pi R^2 \langle q \rangle}$, we obtain:

$$S = \frac{12|e|v_f N_0 (L^2 + R^2) \langle \rho_f \rangle}{(1-K) I_T (1-\beta) \langle \mu_f \rangle} - \frac{K}{(1-K)} S_{A,0} \quad (26)$$

Equations (24) and (26) give different variants of calculation of the integral sputtering coefficient on the basis of experimental data.

Using the assumptions adopted for writing the equation (25), we find the relation of the partial sputtering coefficients of components ($S_{B,p} = S_B c_{B,\delta}$, $S_{A,p} = S_A c_{A,\delta}$) with the concentration of the components in the centre of the film of unit area ($c_{B,f}$, $c_{A,f}$), condensed from the two-component flow and located on the axis of the target at the distance $L$ from its surface. From determining of the component B concentration in the film of unit area, follows:

$$c_{B,f} = \frac{\pi \langle \mu_f \rangle J_i S_B c_{B,\delta} \iint dF \int \frac{d\varpi}{2\pi} \int \cos\alpha \, d\alpha \int dt}{2 \langle \rho_f \rangle h N_0}, \quad (27)$$

where α - the angle between the normal to the target and the line connecting the area element of the target $dF$ with the center of the film.

Hence, after integration we obtain:

$$S_B c_{B,\delta} = \frac{12|e|(L^2 + R^2) \langle \rho_f \rangle N_0 v_f}{(1-K)(1-\beta) I_T \langle \mu_f \rangle} c_{B,f} \quad (28)$$

Respectively, for component A, we have:

$$S_{A,0} c_{A,\delta} = \frac{12|e|(L^2 + R^2) \langle \rho_f \rangle N_0 v_f}{(1-K)(1-\beta) I_T \langle \mu_f \rangle} c_{A,f} - \frac{KS_{A,0}}{1-K} \quad (29)$$

Above was evaluated the filling factor of the target surface by micro-droplets ($K$) and the effective thickness of the droplet phase most difficult to measure. From (29), taking into consideration (19) and neglecting the reflection of the ions, we obtain an independent expression to clarify the magnitude *K*.

$$K = \frac{12/e/\gamma(L^2+R^2)\langle\rho_f\rangle N_0 v_f}{(1-\beta)I_T\langle\mu_f\rangle(\gamma S_{A,0}-1)}c_{A,f} - \frac{1}{\gamma S_{A,0}-1} \qquad (30)$$

### 3. FEATURES OF THE FORMATION OF MONO-COMPONENT FILMS

Describing the essence of the proposed method, it was stated that the change in the charge, energy and component composition of the initial ion plasma flow with a conductive target, which electric potential is regulated within the prescribed limits, and the target itself can be located in a magnetic field of a given configuration. Obvious that the development of a new method should lead to an improvement of some physical and technological characteristics of the films formed.

One of the urgent tasks is to solve the problem of stability of the films. Priori clear that the realization of this problem is associated with the suppression of diffusion processes in the formed film systems after the completion of the condensation process. There are two possible directions: 1-eliminating the driving force of diffusion; 2 – eliminating the transport link of diffusion (under "normal" conditions these are vacancies ).

In this section we present the experimental results, which indicate that the second direction can be realized under the formation of epitaxial films of niobium from the atomic flux in a certain energy range.

In the manufacture of the cathode and the target of the same material, increasing the accelerating potential of the target you can set it self-sputtering regime, which is implemented by us for niobium [2]. Experimental scheme is shown in Figure 1. Conducting target should be viewed as a probe placed in a vacuum-arc plasma, which is able to generate streams of charged particles, depending on the electric potential.

In the absence of a special form magnetic field near the surface of the target, energy distribution of the electrons in the stream, which is distributed from the target toward the substrate, has two peak maximums. Scattered electrons, emitted by the cathode of a vacuum-arc device, have an average energy of 3 eV .Secondary electrons, emitted by the target, are accelerated in its electric field, that is why their average energy is determined by the target's potential .Secondary electrons, reaching the dielectric substrate, will create on its surface negative potential, which will be close to the targets potential (at least at the initial stage of growth). In the presence of ionic component in the condensing flow, the processes of nucleation and growth of the film will take place in the bombardment conditions of the substrate's surface by the accelerated ions. It promotes the formation of an extended transition region (pseudo-diffusion zone) on the border of film-substrate [14, 15, 16]. When placing the target (or substrate) in a magnetic field of special configuration (e.g. such as arched), you can change the exposure of the substrate surface by the flux of fast electrons. Intensity of the flux depends on the degree of permeability of the magnetic field in the space between the target and the substrate for the electrons of given energy.

Decrease in the accelerating potential leads to a reduction in the intensity of an irradiation of the substrate surface by the flux of fast electrons, and simultaneously to decrease in the share of sputtered atoms. Known that the reflected particles are also present in the resulting flow, in addition to the sputtered atoms. We want to highlight the possibility of generating an atomic flux by the target in the mode of reflection of high-energy particles from the target with a zero or a small accelerating potential. When the cathodes of vacuum-arc devices made of refractory metals (Ta, Nb, Zr, Mo, W) are used, the average kinetic energy of ions in the erosive flow lies within (100-150) eV [10]. In this energy range the self-sputtering coefficient of the target is much smaller than one. In the paper [17] was established that the probability of reflection of fast atoms, which interact with the target, reaches a noticeable value in the energy range of (30-200) eV. The reflection coefficient depends on a given element and a particle energy. The reader can find the studies of the processes of reflection in papers [18-20], apart from the fundamental work [17]. Hence, during the interaction of the ion-plasma stream of given substance with the target, for the incoming particles exists a definite value of the energy at which the reflection coefficient reaches its maximum. Therefore, to obtain films in the reflection regime with a maximum deposition rate, target's potential must be changed within a small range near to zero value. It is obvious that the condensing stream has a higher purity compared with the initial stream, since an additional cleaning of the cathode material from the dissolved in it gasses and other impurities occurs during the reflection, due to the dependence of the reflection coefficient not only on the energy of incoming particles but also on their type. This is especially important for active metals (metals- getters).

Typical distribution functions on the energies of niobium particles, generated by the target, are obtained by differentiating the current-voltage characteristics of a multi-electrode probe, are shown in Fig. 2.

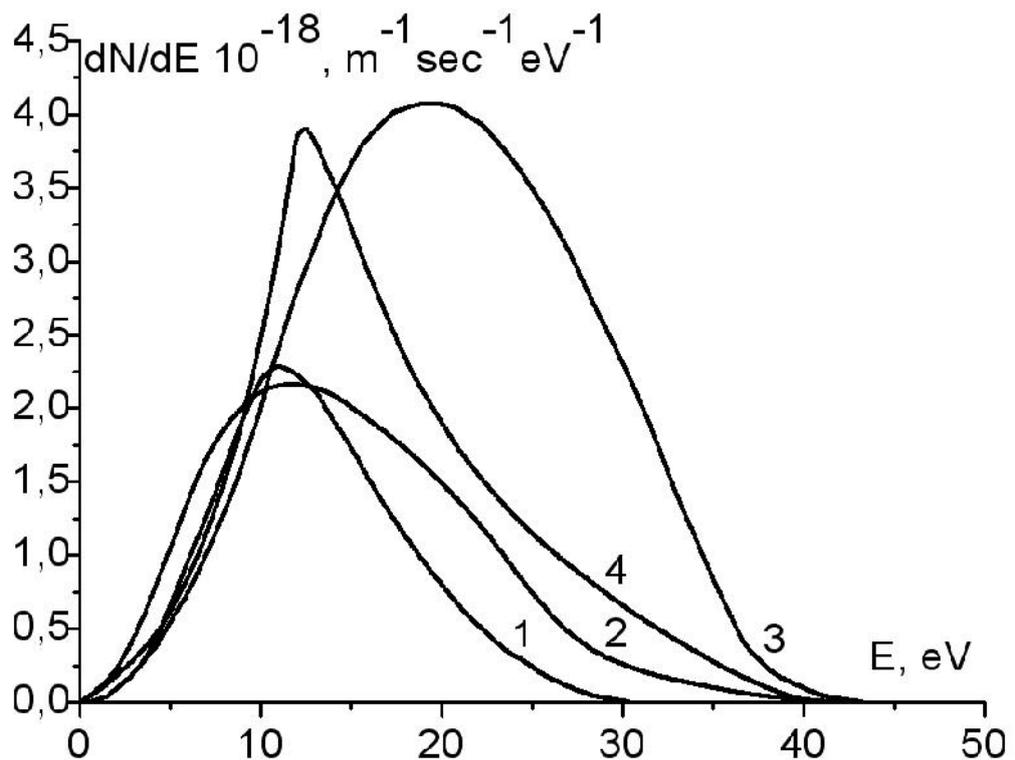

Fig.2. The energy distribution of particles in the transformed flux. 1 - $\varphi = 10^0$, 2 - $\varphi = 20^0$, 3 - $\varphi = 45^0$ at $U = 0$ V; 4 - $\varphi = 20^0$, $U = -20$ V

Presented graphical dependences clearly demonstrate the changes in the energy spectrum when applying a small accelerating potential. When increasing the accelerating potential of the target, the relative contribution of the reflected flux in the total flow decreases, and a function of the atomic distribution over energies tends to the form referred to the sputtered flux, described in the literature, see e.g. [21].

The degree of ionization of the reflected flux depends on the direction of propagation of particles and changes from $\gamma \approx 0{,}2$ when $\varphi = 0^0$, to $\gamma \approx 0{,}5$, when $\varphi = 40^0$, while the reflection coefficient reaches 0,3. Ionization of niobium atoms, generated by the target, is a consequence of the existence of crossed plasma flows of an arc discharge and the reflected flux of the substance (see Fig. 1).

We especially emphasize on the particular features of the erosion process of molecular crystals, during their low energy bombardment, on example of the fullerene target. The theory, adequately describing the collective interactions of the particles, currently does not exist. The paper [12] presents a method of estimating of critical values of ion's energy, below which the multi-particle effects prevail. According to this method, as a criterion of the multi-particle interaction can be selected the energy value of the incident particle, when its speed will not exceed the speed of the sound in the target's material. In the case of low-energy irradiation of fullerene by ions of this type, using the criteria of multi-particle interaction for the fullerene molecule, we can estimate the critical energy of bombarding particles, below which the momentum of the incident ion $C_{60}$ will be taken as a whole. In this case the energy of the incident particle can go on increasing the kinetic energy of molecular motion and their internal energy. There is possible exit of the molecules $C_{60}$ from the target due to the physical molecular sputtering and due to the radiation-accelerated sublimation of the molecules [22].

As it is indicated by the experimental data of author [23], in the films condensed from the flow of particles with the energy not more than 100 eV, the compressive stresses of structural origin appear, which level is determined by kinetic and thermodynamic parameters. This indicates on the formation in the films of secondary condensation defects of an interstitial type.

We try to address the problem of suppression of diffusion processes in the formed film systems after the completion of the condensation process by eliminating the transport link of the diffusion in the formed films. Most promising is the energy range corresponding to the curve 3 on Fig.2. This follows from the following considerations.

Obvious that the transfer of the momentum from the incident particles of high energy under certain conditions, to the inside of the interior of the crystal lattice, can cause migration of vacancies to the surface of growth (ordering occurs in the structure of growing in vacuum crystal). This shock-activated ordering should occur at maximum during condensation of a mono-energetic beam of atoms and ions, whose kinetic energy is close but does not exceed the threshold displacement energy of atoms in the lattice ($E_d$). For niobium $E_d$ =28-36 eV. In the real conditions the condensable particles always have some energy distribution, within which the condition $\langle E \rangle \leq E_d$ can be satisfied, where $\langle E \rangle$ - an average energy of the particles in the flow.

This raises the question of the boundary values of the energy spectrum. Minimum energy value should be chosen from the condition if the activation of process of vacancies' concentration reduction (including the directional migration of vacancies to the free film surface), and maximum energy value – from the condition of generation of minimum number of stable Frenckle pairs with one high-energy condensing particle. Energy, which had been transferred to the atom of the lattice in the binary elastic collisions by incident particle, varies from small values during sliding collision up to a maximum value in frontal collision $E_{max} = E^0$, where $E^0$ - the initial energy of the incident particle. Therefore when $E_d < E \leq 2 E_d$, a single incident particle is capable of creating no more than one Frenckle pair, and the vacancies will be concentrated near the free film surface in the area of their formation. Intrinsic interstitial atoms, which momentum is directed deep into the film, can penetrate deep into the crystal lattice mainly due to the chains of focused collisions, so as with a decrease in energy of the incident particles the probability of channeling is significantly decreases [24].

When $E_0 < E < E_r$(hkl), ($E_r$ - substitution energy, sufficient to form a chain of substitutions of the atoms along the direction hkl, and $E_0$ – the lower limit, at which the chain is no longer existing) and the chains of collisions transfer only energy (region where focusons exist), then atoms can occupy the free positions in the chain of collisions, beginning of which comes from the near-surface atoms. This corresponds to the displacement of a vacancy by one position to the beginning of the chain. $E_0$ for most solids is about 1 eV [15] and in order of magnitude it is not less than the migration energy of vacancies. Provided that the energy of the condensing particles lies in the range $E_0 < E \leq 2E_d$, $\langle E \rangle \sim E_d$, we should expect that in the film a non-equilibrium concentration of vacancies will be reduced by many orders of magnitude. That will lead to a practical destruction of the transport link of diffusion in the film samples after their formation. The assessment of reduction of the vacancies concentration in the process of film formation can not be made analytically, the real way to do this – computer experiment.

When using such an energy interval, the probability of radiation damage in the formed crystal is negligible. Fast atoms, falling on the film surface, are able to penetrate a small depth into the film, while part of taken roots atoms diffuses in the crystal lattice, saturating it with own interstitial atoms. The "subsurface" mechanism of growth is predominantly realized during the process of condensation, as most of the atoms embedded in their equilibrium positions when entering the free surface of the film volume [23].

The key findings of the mono-component films, which were obtained when using the energy distribution of condensed particles, are shown in the curve dependence 3 Fig. 2, this satisfies the given above energy interval for niobium.

The experimental results of film formation from fluxes of high-energy particles give the grounds for assumption that the re-crystallization in the condensed films occurs under the interstitial diffusion mechanism .Found that the films, formed in the above energy range, are characterized by a high speed of re-crystallization, beginning from the earliest stages of the growth, and a high level of the compressive macro stresses, unlike the films deposited by the thermal evaporation [23].

It is known that during the thermal deposition on mica such substances as V, Nb, Ta, the nucleation and growth of the epitaxial films occur in at almost equivalent three azimuthal orientations [25]. Experience shows that the epitaxial niobium films on mica, formed from a stream of particles of high energy, also have a three-orientation origin. Therefore the film of niobium which is not thicker than 3 nm, formed on an artificial mica- fluorine-phlogopite at a substrate temperature $T_S = 700^0C$ in the energy range $E_0 < E \leq 2E_d$, is continuous epitaxial, and its grains are in the tree azimuthal orientations. However, at this stage of growth, the distribution of azimuthal orientations is not equiprobable. The film with thickness of 15,6 nm, formed at the same temperature, is predominantly a single-orientational, as it follows from the electron diffraction pattern shown in paper [23]. The transition temperature of the film into the superconducting stage is 8,35 K, when $\lambda = R_{300K} / R_{10K} = 6,5$. Sufficiently high value of for the film of given thickness, at the same conditions, indicates a low scattering power of the carriers by the film surface, which is indirectly confirms the high smoothness of the film surface.

As it was established in paper [26], an elastic macro deformation of compression of thin niobium films, not separated from the substrate, is anisotropic in the plane of the film at the thickness less than 40 nm. With increasing thickness of the condensate, an anisotropy degenerates (Fig. 3).

In the early stages of growth under the anisotropy macro deformations conditions, and when the macro deformations caused by the anisotropy of the epitaxial deformations along the directions [001] and [1$\bar{1}$0], a collective re- crystallization occurs. Under the collective re-crystallization the vector sum of lateral diffusion fluxes of an intrinsic interstitial atoms (IIA) leads to the preferential growth of the same orientation grains. Grains in the other orientation occupy a small area in the matrix.

In the energy range $E_0 < E \leq 2E_d$, when the temperature of the substrate changes from room to 1/3 $T_m$, a polycrystalline and epitaxial films form. While in the crystal lattice an increased level of contraction macro stresses of structural origin is implemented. The occurrence of macro stresses is associated with the fixation in the film of secondary condensation implantation defects, which concentration is determined by the law which regulates a temperature change of the film during its formation [26]. Introduction of the IIA in the crystal lattice and decrease in volume concentration of vacancies causes the long-range effect. During the condensation of homo-epitaxial film of niobium with a thickness of about 500 nm on the side of niobium disc of 30 mm in diameter and a thickness of 3 mm, experimentally established that the region with an increased level of concentration macro stresses exceeds the thickness of the deposited film (Fig. 4) [27]. The interstitials, diffusing during the condensation process from the film into the volume of the sample, recombine with the vacancies. In this processes also take part focusons and dynamic crowdions [15], while the concentration of vacancies significantly decreases. This leads to destruction of the transport link of diffusion in the sample after film deposition. The measured rate of oxidation of the niobium disc with one-sided coating decreased by more than 5 times compared to the initial sample without coating. It formed the basis of our proposed method of protection of bulk materials from corrosion [28].

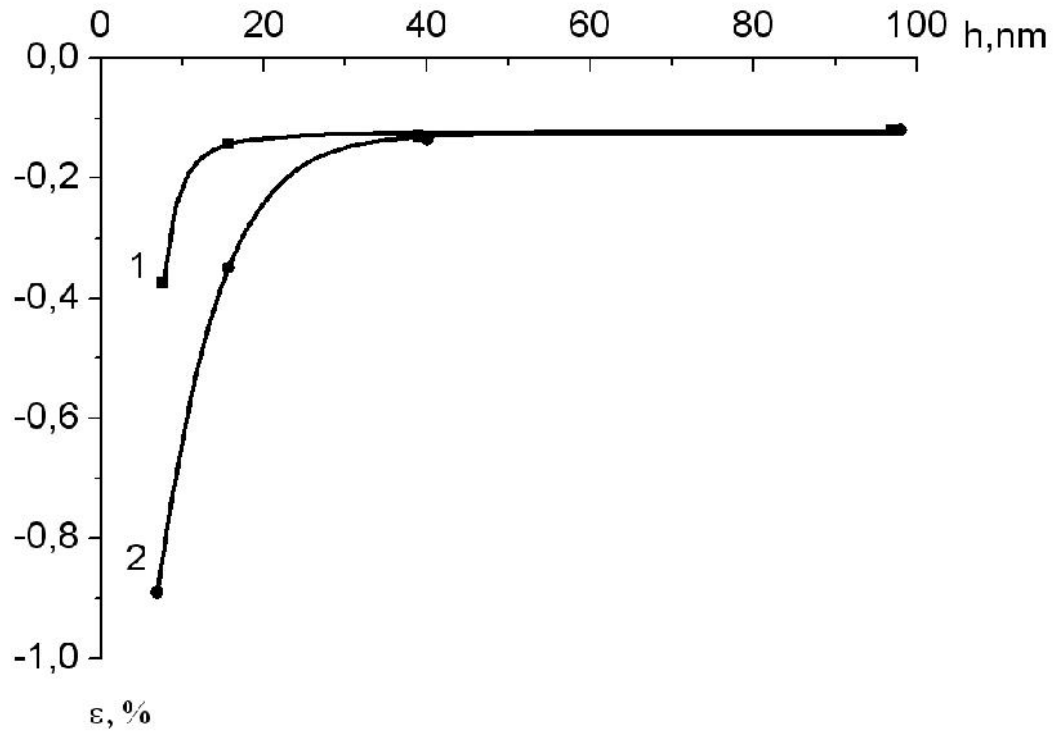

Fig.3. The dependence of macrodeformation on the thickness of nonseparated from mica substrates Nb films for two directions in the basic plane.

Established that during stepwise reduction of the substrate temperature from $1/3\ T_m$ to the room temperature in the process of condensation of niobium atoms in the energy range $E_0 < E \leq 2E_d$, we can create some layers in the film with different level of compression macro deformations. The presence of layers with high and low level of deformations is easily detected

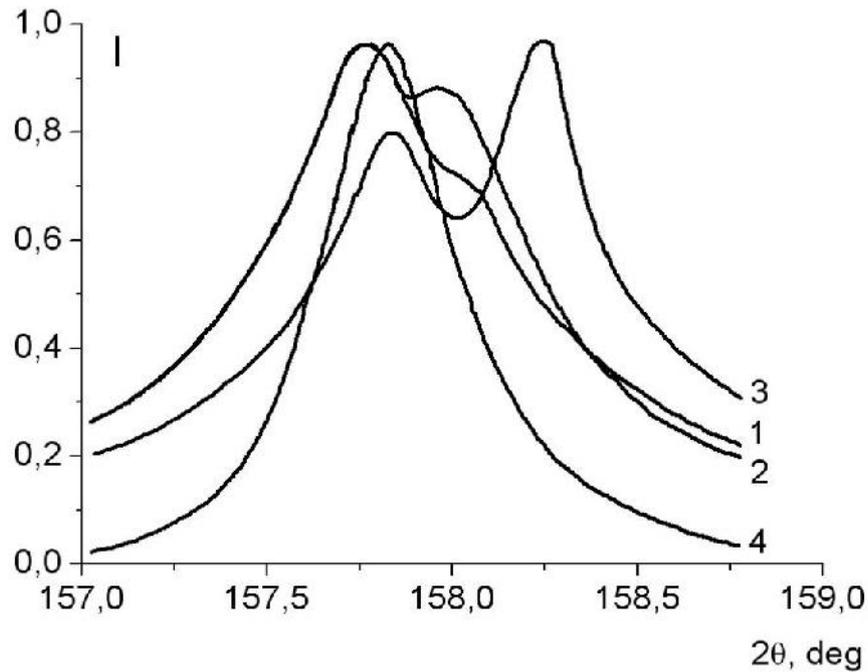

Fig.4. The redistribution of intensity in the maximum of $K_{\alpha 1}$ component of (220) reflection from homoepitaxial system film – massive Nb substrate at various glancing angle ($\beta$) of primary beam. 1 $\beta_1 = 10^0$, 2 – $\beta_2 = 7^0$, 3 – $\beta_3 = 5^0$ (front) 4 - $\beta_4 = 5^0$ (back)

by X-ray diffract metric studies due to the separation of Bragg's reflections (Fig. 5) [26].

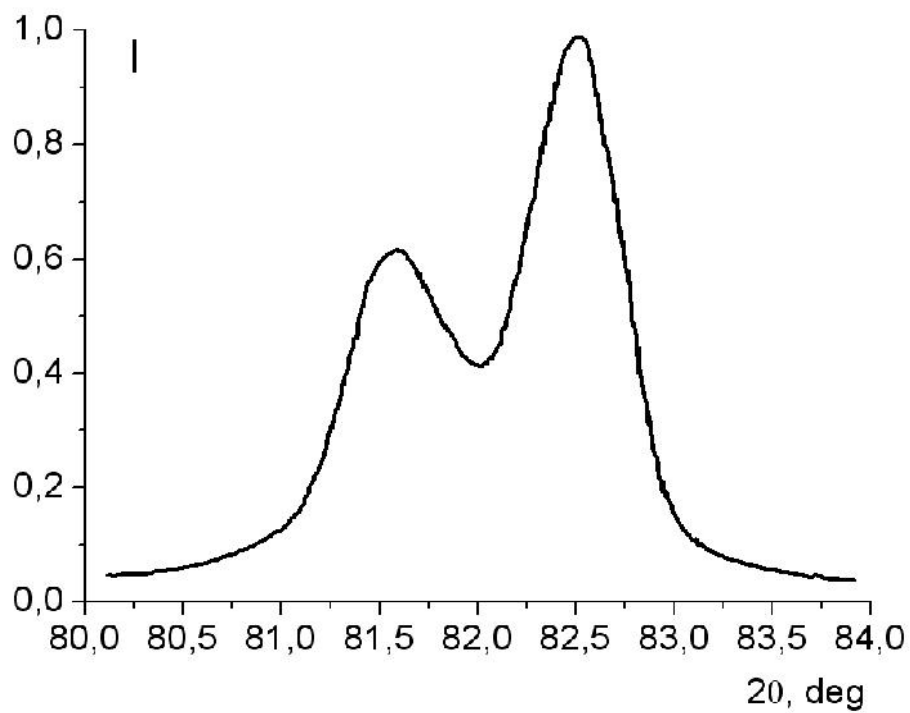

Fig.5. The splitting of (220) Bragg reflection from Nb film, formed at step change of substrate temperature. $T_{S1} = 700^0$ C, $h_1 = 40$ nm; $T_{S2} = 50^0$ C, $h_2 = 100$ nm.

Annealing for several hours at the temperature of formation of the lower layer ($T \approx 1/3\ T_m$) leads to the disappearance of the separation of Bragg's peaks and reduction in the level of macro deformations. Macro deformations remain substantially higher than the deformations that are fixed in the films formed under isothermal conditions at $1/3\ T_m$. Reduction of non-equilibrium vacancy concentration in the films after their formation in the energy interval $E_0 < E \leq 2E_d$, leads to a number of features not only in the structure, but also in the physical properties of the condensed films. In the films of niobium on silicon substrates, formed in the above energy range, when using the method of diffuse saturation of the isthmus by substrate's material, we failed to form a weak link for the Josephsons junctions. Technique of saturation of the niobium crystal lattice by silicon is based on local heating of the isthmus by electric current, passing through it. However in the films, formed in the energy range $E_0 < E \leq 2E_d$, during the transition from the standard regimes to the regime with increased current density, a smooth change in the concentration of silicon in the film was not implemented, because when we got a certain critical current density, the isthmus was destroyed. Thin niobium films, formed in the above energy range, also revealed a high temporal stability in relation to the not controllable atmospheric impact. For example, the free niobium film with thickness of 15,6 nm, kept without special measures on a copper mantle over 20 years, shows no signs of degradation (see Fig. 6). The data of electron microscopic studies of the same film, which image is given in paper [23], was used as the initial.

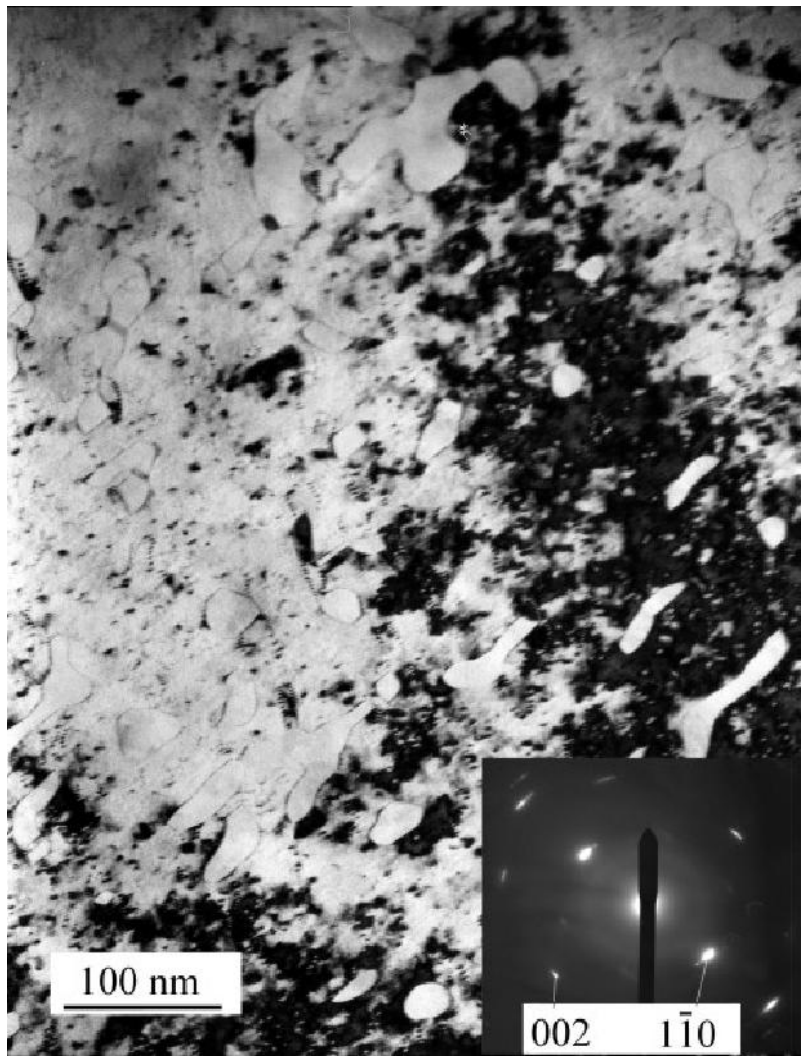

Fig.6. The recrystallization in the Nb thin layers on the mica. $T_S = 700^0\ C$, $h = 15{,}6$ nm, $p = 7\ 10^{-5}$ Pa. TEM image of separated film after more then of 20 years storage on the cooper grid.

The set of an experimental results given above, which describes the film growth in the energy range $E_0 < E \leq 2E_d$, indicates the occurrence in them of interstitial mechanism type of diffusion. To find the general laws of diffusion phenomena in the process of film growth from the particle flux with thermal and high energy, we can use data from the following papers [29, 30]. The papers indicate an acceleration of the intra-granular and grain boundary diffusion in the process of condensation (the effect of condensation-induced diffusion during the thermal deposition of the films). Obvious that the excess vacancies can affect the diffusion processes in the case when their super saturation artificially pegged. This is possible only in the dynamic mode of nucleation and disappearance of vacancies under a self-consistent actions of their sources and drains [31]. In the process of film growth from the thermal fluxes, the source of excess vacancies is its surface. Vacancies can diffuse into the grains and to their boundaries. Combining this with the results, obtained by the author, you can make general conclusion, that the effect of increasing the diffusion rate is proportional to the excess of the non-equilibrium concentration of point defects during the film growth. This effect is universal and applicable to the vacancy and interstitial diffusion mechanisms. Essence of condensation-induced diffusion is in the acceleration of the bulk and grain-boundary diffusion in the condensation process. This happens as a result of increased concentration in the lattice of point defects of vacancy or interstitial type, generated mainly near the surface of the growth. The dynamic equilibrium of the defect flow in the crystal lattice is ensured by both internal and external drains in the film, as well as by occurrence of reactions between the point defects. If the energy of deposited particles will be in the range $E \gtrsim E_0$, $\langle E \rangle \sim kT$, then the vacancy mechanism of condensation-induced diffusion will be realized. When $E_{i,f} < E \leq 2E_d$, $\langle E \rangle \sim E_d$, where $E_{i,f}$ - the formation energy of interstitial atom, an interstitial mechanism had been observed. Typical values of the formation energy of interstitial atoms lie within (2-5) eV [32, 33]. Energy range of condensing particles within which interstitial diffusion mechanism is realized, can also be expressed through the energy of sublimation of the substance per atom ($E_S$): $E_S < E < 10E_S$, $\langle E \rangle \sim$ (4-5) $E_S$.

With increasing energy of condensing particles the contribution of collision cascades in the process of generation of point defects in the lattice is also increases. While the role of radiation-induced diffusion in the condensed film also increases. The main difference between the contribution of radiation-induced diffusion from the condensation-induced diffusion is as follows; in the first case, the source of point defects (a cascade of collisions) generates both vacancies as well as IIA. In the second case we can choose such energy intervals of condensable particles, under which either vacancies or IIA are mainly generated.

Diffusion fluxes, leading in the condensation process to the saturation of grain boundaries by point defects, reduce their resistance. This is true in the general case of condensation-induced diffusion. However, its speed is different during the occurrence of diffusion by the vacancy and interstitial mechanism. It is worth to emphasize that the difference in the rate of re-crystallization at the initial stage of formation of a continuous film. Niobium films, condensed during the thermal deposition have more pronounced dip in the profile in the area of grain boundaries, compare to the films, formed from the particle' fluxes in the energy range $E_0 < E \leq 2E_d$, $\langle E \rangle \sim E_d$) at ceteris parabus. And the boundaries contain an increased, compare to the volume of the grain, concentration of impure atoms, entering the film from the substrate and the residual atmosphere of the vacuum chamber. Depth of the dip determines the height of the potential re-crystallization barrier. Qualitatively we can state the following during the formation of the films from high-energy particle flux, small island have flattened, but not domed shape [14]. During the formation of a continuous film, the own interstitial atoms diffuse to the inter - grain boundaries, saturating the borders and aligning their profile, reduce the potential barriers, which can not be said about vacancies. It should be noted, that during the thermal deposition the concentration of impurity at the boundaries of the film increases during their re-crystallization. Impurity, carried away by the front of the re-crystallization, eventually reduces the speed of the re-crystallization.

On the contrary, the compressive stresses, arising within the grain during deposition of high energy particles, are the barrier, which prevents the implantation of impurity atoms in a crystal lattice. Indirect evidence of purity of the niobium films with thickness h = 100 nm is their superconducting transition temperature $T_C$ = 9.15 K and the ratio of the sample's resistance at temperatures 300 K and 10 K $\lambda = R_{300K}/R_{10K}$ = 41 [23].

There is still no consistent theory of nucleation and growth of films from a stream of accelerated particles, due to the complexity of the processes, which occur during film condensation. Deposition of the films from atomic fluxes is a non-equilibrium process in an open system, where an exchange of flows of matter and energy with the environment is happening. While in the process of growth the formed structure loses its stability and irreversibly transforms into a non-uniform stationary state. Explanation to the formation of the films should be made from a position of self-organization theory, which studies the general laws of formation, sustainability and destruction of the temporal and spatial structures in the complex non-equilibrium systems of different nature [34, 35]. Let us consider an example of manifestation of self-organization process of secondary condensation defects of interstitial type in the niobium films, causing compressive structural macro deformations. We should follow the evolution of structural macro deformations (and associated macro stresses) in epitaxial films of niobium, formed on fluorine- phlogopite (fph) substrates in the energy range $E_0 < E \leq 2E_d$ when $T_S \approx 1/3$ $T_m$ ($\approx 700^0$ C) in their growth. As stated above, the films of Nb with thickness no more than 3 nm are continuous, epitaxial, fine - grained, with the grains located in three almost equivalent azimuthal orientations, i.e. nucleation and growth of the films occur on island mechanism. During the film formation we should separate macro deformations of structural origin and compressive isotropic thermal macro deformations in the film plane, due to the difference in the thermal expansion coefficients (CTE) of film and substrate. Temperature dependence of macro deformations of niobium films, formed on fluorine-phlogopite, shown on Fig. 7 [26].

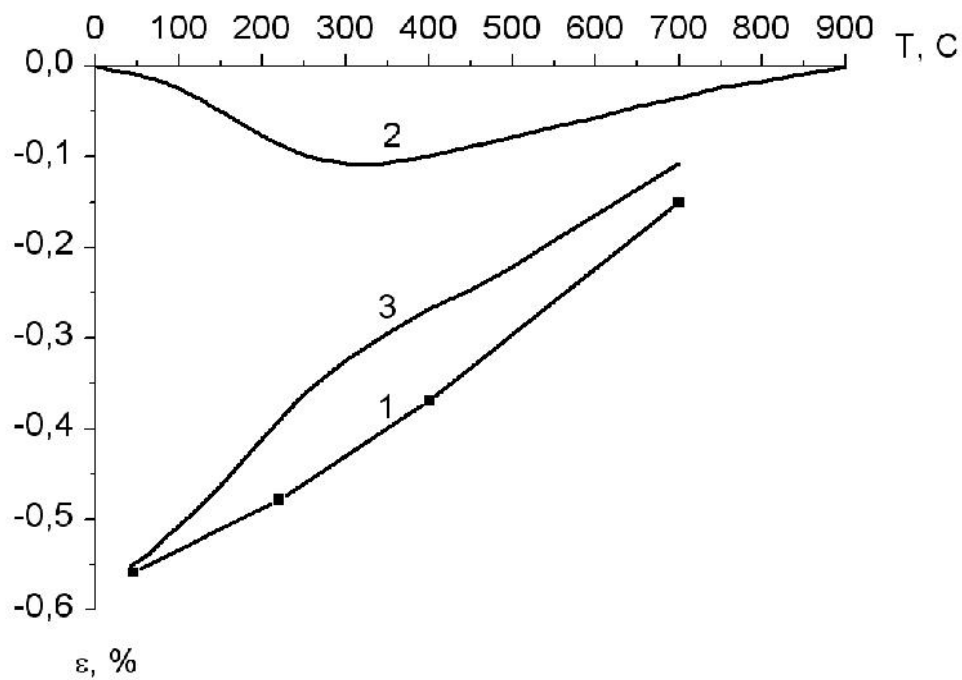

Fig.7. The macrodeformations temperature dependence of the nonseparated from mica substrate Nb films ($h \approx 100$ nm). 1- the experimental data of resultant macrodeformation, 2 – the thermal macrodeformation, calculated from the Nb and mica coefficient of thermal expansion difference, 3 – the structural macrodeformation, obtained by subtraction of curves 1 and 2.

Within the grain, from earliest stage of film growth, the elastic anisotropic compression deformation of structural origin is realized. The occurrence of this deformation has a dual nature. An anisotropic deformation, deformation in the direction [001] and [1$\bar{1}$0], IIA epitaxial deformation and deformation, caused by the formation of own interstitial atomic complexes bring contribution to the total deformation. Due to the fact that the inconsistencies in the substrate lattices and the film differ in magnitude and sigh [26], distribution of density of the OIA complexes in the plane of crystal lattice is described by a wave-like function. This wave-function visualizes by electron-microscopic images of the formed films in the form of quasi-periodic contrast, which average period varies with the film thickness [23]. Based on the fact that the phase contrast of quasi-periodic formations corresponds to the dislocation contrast, in the field of epitaxial stresses IIA combine into dislocation implantation loops. The arrangement of the loops leads to a decrease in energy of the deformed lattice. Probability of the association of the intrinsic interstitial atoms into complexes and their location in the lattice changes during the film growth. At the early stage IIA mainly diffuse to the free film surface and to the grain boundaries, which facilitates rapid re-crystallization process. At this multistage phase of the film growth a certain nonlinear wave process of rising and relaxation of compressive deformation within the grain can be observed, due to the occurrence of diffusion processes in inhomogeneous stress field. The maximum value of deformations and effective linear grain size are in the causation. Self-consistent process of re-crystallization and change in size of deformation end with relatively stable phase in the film formation (inhomogeneous stationary state), when the effective linear size of the grain in the plane much more than grain's thickness (practically it's the realization of one of the azimuthal orientation of the film). Saturation of the crystal lattice of IIA, reacting with each other to form complexes with varying number of particles, leads (with increasing thickness of the film) to the isotropy of macro deformation. The isotropy of macro deformation tends to its asymptotic value, which depends on the substrate temperature (Fig. 3). While macro stresses remain anisotropic due to the anisotropy of modulus of elasticity in the plane (110) [26].

After establishing of stationary level of isotropic macro deformations, formed in the plane of the film under $T_S$ = 700$^0$ C, this corresponds to the value ε = 0,13 % on Fig.3, additional application of thin layer of niobium (without opening the vacuum chamber) with thickness 3 nm at room temperature, which at small thickness does not cause separation of the Bragg's reflections (see Fig. 5) causes the anisotropy of macro deformations in the formed films. And the inequality of deformations along the directions [1$\bar{1}$0] and [001] changes its sign [26]. The contrast, associated with the deformation field, which is directly related to the distribution in the crystal lattice of the IIA complexes (Fig. 8) appears more clearly (compared with the isothermal deposition [23] ) on the electron-microscopic images of niobium film).

Association reaction of IIA of niobium into the dislocation implantation loops with increasing thickness of the film occurs in the field of contracting macro-stresses, which level in the film plane grows with decreasing temperature. Therefore, as it follows from the initiated by stresses mechanism of predominant nucleation (SIPN) [36], orientation of the implantation loops becomes energetically advantageous in the planes which are parallel to the basal plane of the film. This fact evidenced by the growth of period of niobium cubic lattice along the direction [110] with a decrease in temperature at constant equal in magnitude values of lattice period along

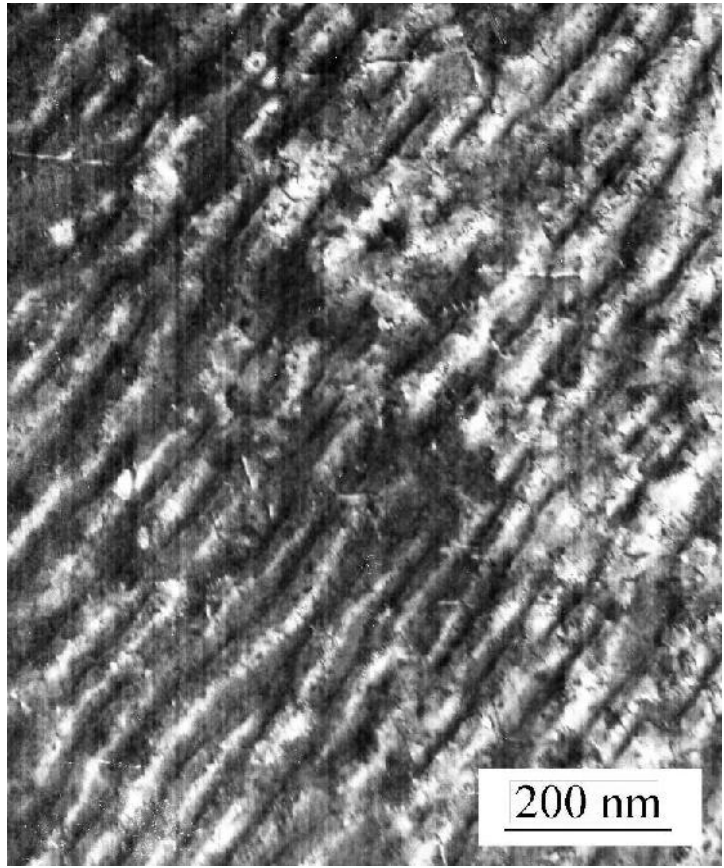

Fig.8 Visualization of the quasiperiodic arrangement of the IIA complexes in the process of self-organization of secondary condensational defects in the niobium films, formed during the stage reduction of the substrate temperature. $T_{S1} = 700^0$ C, $h_1$ =80 nm, $T_{S2}$ =50$^0$ C, $h_2$=3 nm.

the directions [1 $\bar{1}$ 0] and [001] (Fig. 9).

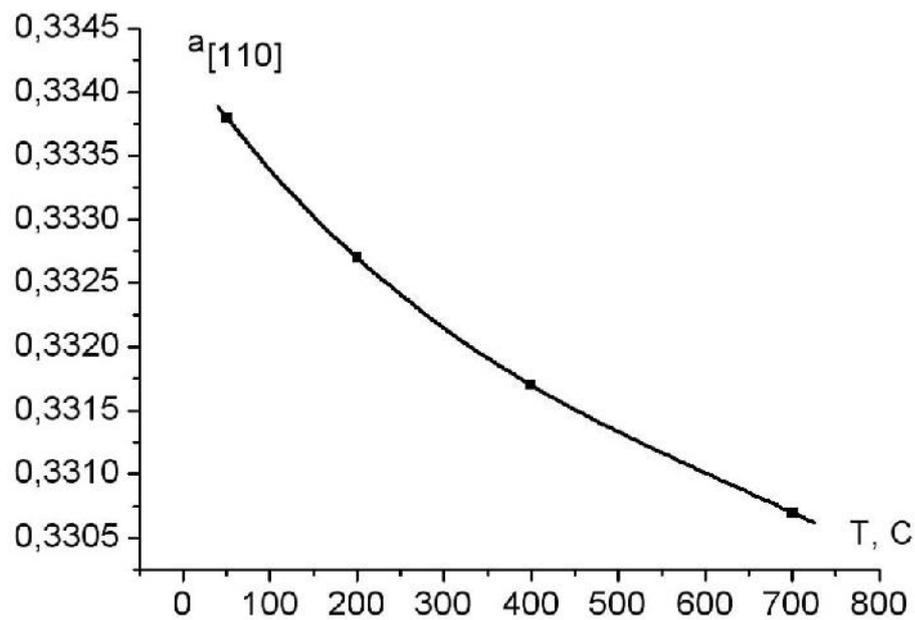

Fig.9. The temperature dependence of the lattice spacing of the Nb films along [110] crystallographic direction, normal to basic plane ($h \approx 100$ nm).

Concentration of the IIA complexes formed in this way is in co relational (self-consistent) relationship with the value of elastic lattice macro-deformation, therefore the value of concentration of the secondary condensational defects and the value of macro-deformation tends to its asymptotic value, corresponding to a given temperature of the film formation. During the film growth while achieving the single-orientation state of the film, a decrease in power output of internal flows had been observed. And fast atoms introduced into the lattice mostly move to the free film surface. Such state of the film formation process can be characterized as a stationary, when growing film is under the impact of small perturbations, generated by the flow of its own atoms, introduced into the lattice. We can identify, for example, micro-deformational lattice perturbations from implanted atoms and perturbations of the vibration spectrum of the film crystal lattice.

The formation of thin niobium films on fluorine phlogopite substrates during deposition of atomic fluxes of thermal energy should be considered with the similar positions. It is necessary to keep in mind that the saturation of the crystal lattice by the excess non-equilibrium concentration of vacancies occurs in the process of growth.

As a result of reaction of vacancies among themselves, vacancy complexes are formed, which area of predominant location inside the grain determined by the desire of the system to reduce the level of deformations in the lattice. In the formed niobium films in the location of vacancy complexes inside the grains with a sufficiently large efficient linear size some quasi-periodic structure also visualized on the electron-microscopic images [37, 38].

## 4. THE ROLE OF KINETIC AND THERMODYNAMIC PARAMETERS IN THE FORMATION OF THE PHASE COMPOSITION OF ALLOY FILMS OF Nb-Sn DURING DISPERTION OF TIN BY NIOBIUM IONS

Niobium-tin alloy films were condensed on fluorine-phlogopite and silicon substrates from a stream of particles, generated by a tin target during its bombardment by niobium ions in the range of accelerating voltages U=(400-3000) V. Samples for research of thickness of condensed films and components concentration were deposited on silicon substrates under the room temperature. Substrates were located along the arc of a circle of radius 200 mm, whose center coincides with the center of the target (see Fig.1). Concentration dependence of the alloy components is determined by X-ray fluorescence analysis method of the intensity of the analytical lines. Calculation of mass fractions of chemical elements Nb and Sn was made by the method of fundamental parameters [39].

The work was not aimed strictly define concentration limits during the formation of a binary niobium-tin alloy. Experimentally confirmed the possibility of controlling the alloy concentration when an energy of sputtering ions and a substrate temperature were changing. For practical purposes we can assume that for the above scheme of obtaining the alloy, mode of sputtering of the tin target by niobium ions is realized when $U \geq 400$V, ($\langle E \rangle \geq 1200$ eV, and $\langle q \rangle = 3|e|$), while the concentration of tin does not exceed 2%. Growth of accelerating potential of up to 3000V causes an increase in the tin concentration in the alloy of up to 65% (Fig.10).

It should be noted that the use of a grounded cylindrical screen, which limits the aperture of the ion-plasma beam and at the same time protects from entering the initial particles stream on substrates (Fig.1), causes a decrease in the degree of ionization of the beam spray. It happens due to the fact that niobium ions reflected from the screen fall on the target and neutralize by interaction with its conductive surface. Value of the neutralized part of the stream depends on the diameter and length of the cylindrical part of the screen. The presence of the neutral component in the initial beam leads to an increase in the surface concentration of niobium and volume concentration of niobium. Because the atoms of niobium, located on the target surface, can be implanted in its volume. The rate of condensation of niobium-tin alloy was determined from experimental data of measuring the thickness of the films. The angular dependence of the rate of condensation at various accelerating potentials of the target presented on Fig.11.

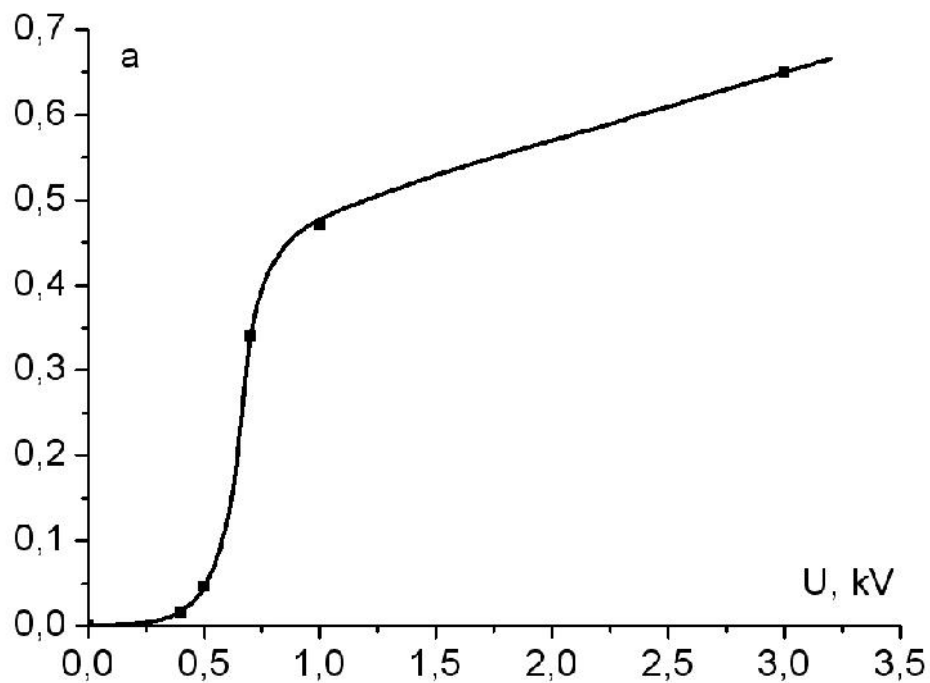

Fig.10 Dependence of the concentration of tin in a spray stream at $\varphi = 0^0$ on the accelerating potential of the target.

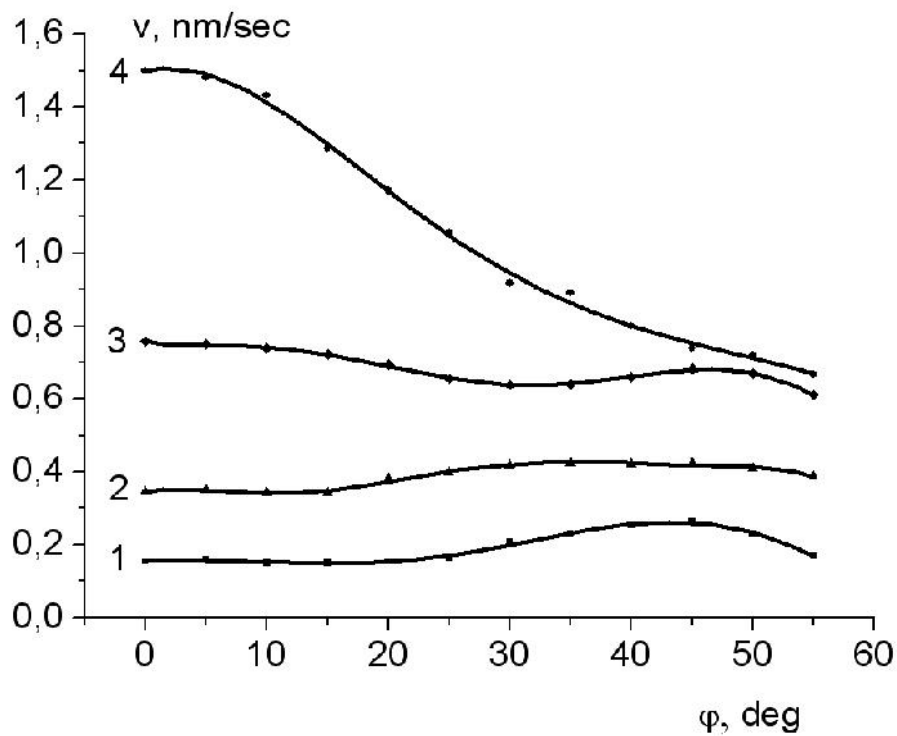

Fig.11. Angular dependence of the condensation rate of the total flux, generated by the tin target, at distance of 200 mm from its center at various accelerating potentials of the target. 1-400V, 2-700V, 3-1000V, 4-3000V.

Fig.12 shows the angular dependence of the concentration of the tin in the films. It is obvious that with increasing the accelerating potential, the normal component of the ion velocity to the target surface also increases. At the same time in the approximation of the angular distribution of the ejected particles, we should expect differences in the output of sputtered particles with increasing an angle of their departure from the target ($\varphi$) when energy of bombarding ions is changing [12]. This conclusion for the integral sputtered flux is confirmed as a whole (see Fig.11).

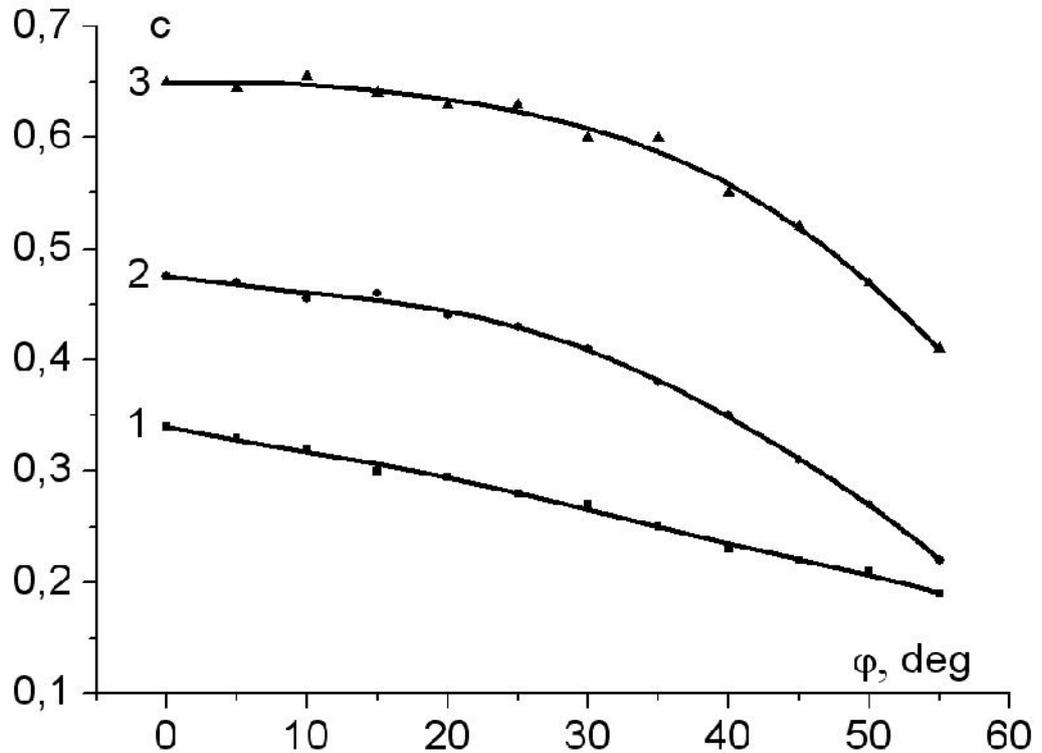

Fig.12. shows the angular dependence of the concentration of the tin in the films at the at various accelerating potentials of the target. 1 - 700 V, 2 – 1000 V, 3 – 3000 V

Angular concentration dependence at different accelerating potentials has some differences, discussions of which will be carried out in the next section. Fig.13-16 show the specific electron diffractions and the typical structures of the formed alloy. The samples for electron microscopic studies were obtained on fluorine-phlogopite substrates in the temperature rate of $200^0$ C – $900^0$ C, when $\varphi = 0$. The use of tantalum substrate-holder allowed to make heating of the samples to the temperature above $1000^0$ C.

It was established that at a substrate temperature above $300^0$ C a decrease in the condensation coefficient of tin is observed. Films obtained at $T_S = 900^0$ C, $U = 3000$V, practically do not contain tin, while at $T_S = 300^0$ C, $U = 3000$V $NbSn_2$ compound is formed (Fig.13).

Nb – Sn alloy films, formed on the f-ph substrates at $T_S < 400^0$ C are characterized by finely dispersed structure. The annealing of such films leads to their re-crystallization. Also was established that heating the sample, which was obtained at $T_S = 200^0$ C, $U = 1000$V, by electron beam in the column of the microscope, causes re-crystallization of the phase, mostly containing $NbSn_2$ compound (Fig.14).

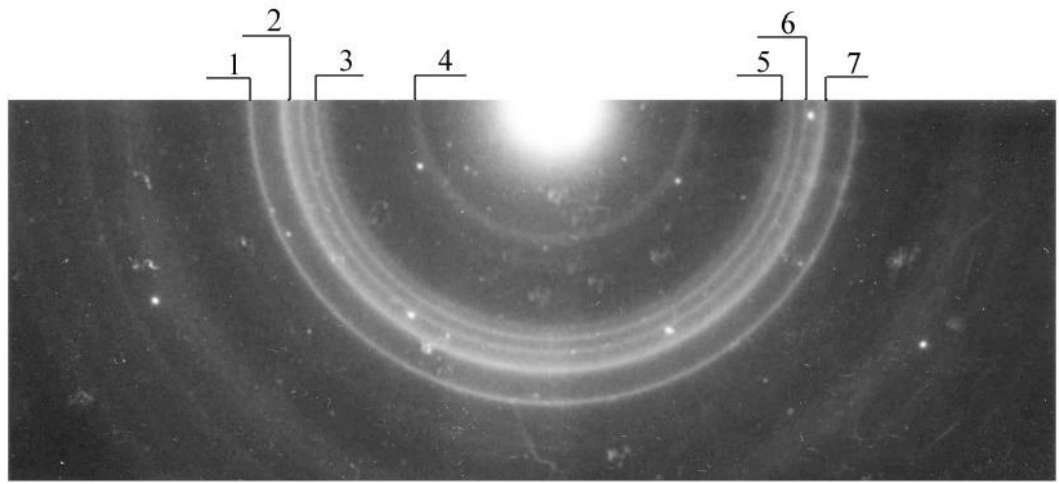

Fig.13. The diffraction pattern of the film with chemical composition corresponded to $NbSn_2$ compound, obtained at $U = 3000$ V, $p = 8 \cdot 10^{-5}$ Pa, $T_S = 300^0$ C. 1 – (402) + (422), 2 – (022),  3 – (220) + (602), 4 – (400) + (111), 5 – (113), 6 – (313), 7 - (800) + (711).

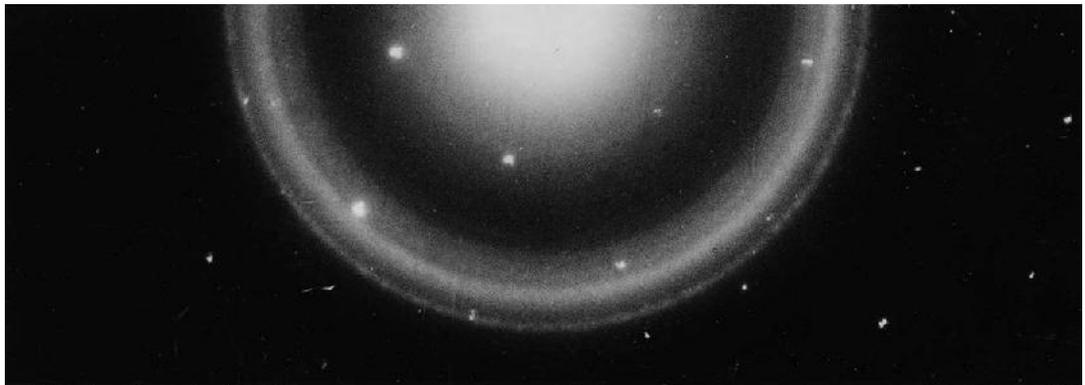

a)

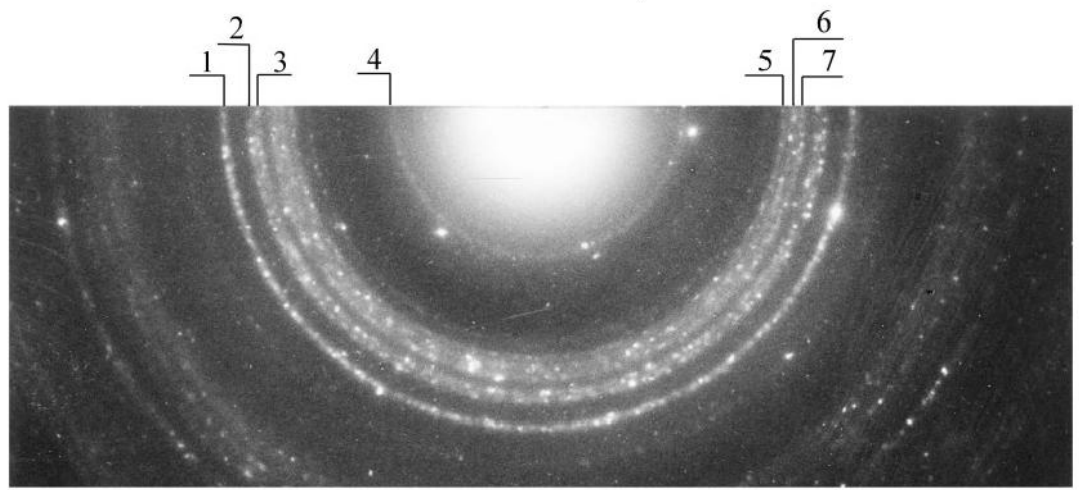

b)

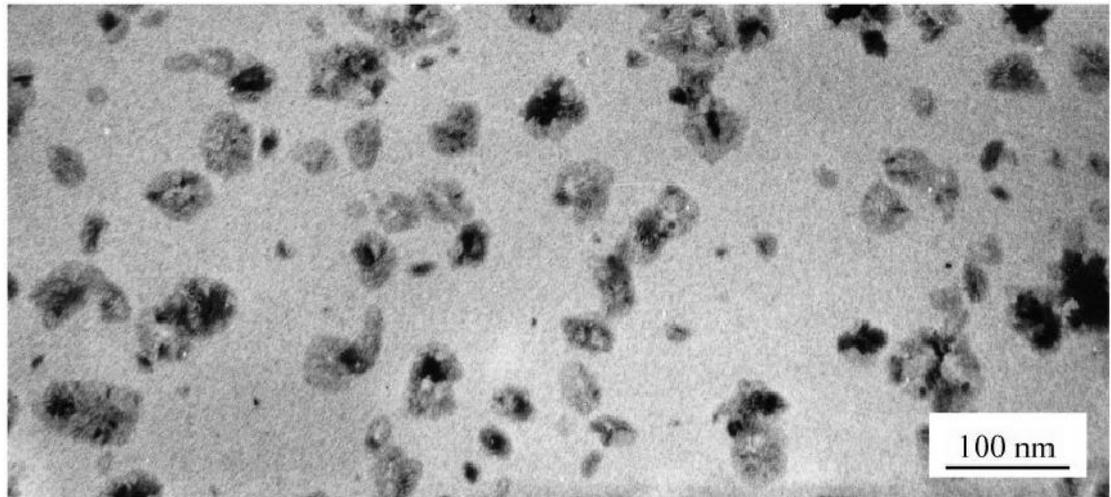
c)

Fig.14. The initial stage of re-crystallization of NbSn$_2$ compound in the film, formed at U = -1000 V, p = 7 10$^{-5}$ Pa, T$_S$ = 200$^0$ C, a) - total diffraction of original film, b) - total diffraction of the film after heating by electron beam in the column of electron microscope, c) - TEM image of the film after heating 1 – (402) + (422), 2 - (800) + (711) 3 – (022), 4 – (400) + (111), 5 – (113), 6 – (220) + (602), 7 – (313).

There is a superconducting compound Nb$_3$Sn and pure niobium in the alloy, obtained at T$_S$ = 300$^0$ C and U = 400V, which was then annealed at 900$^0$ C for 60 sec. (Fig.15).

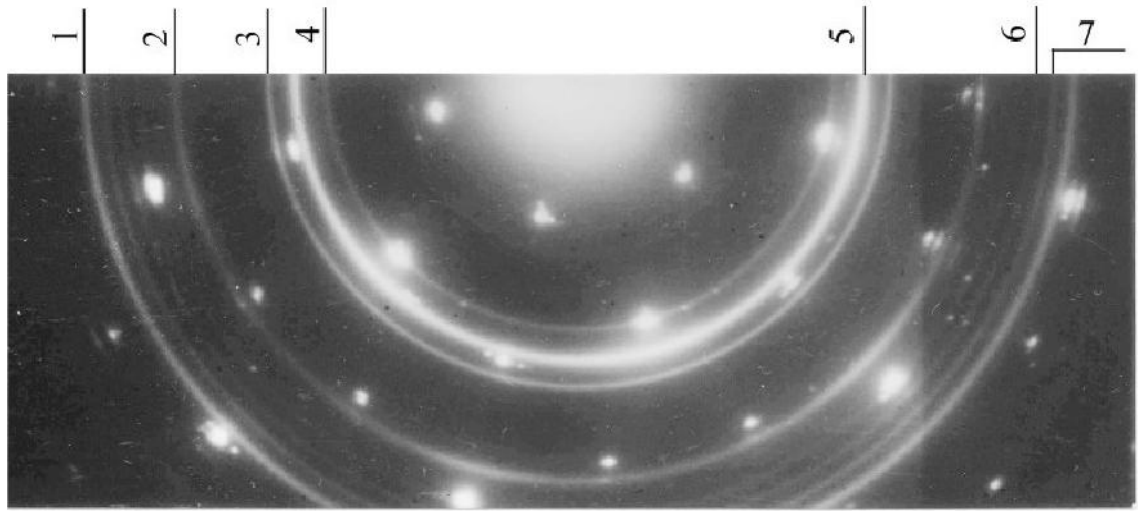

Fig.15. Phase composition of Nb-Sn alloy, formed from the stream of sputtered atoms at U = - 400 V, p = 5 10$^{-5}$ Pa, T$_S$ = 300$^0$ C, followed by annealing in the vacuum chamber at T$_S$ = 900$^0$ C, p = 3 10$^{-5}$ Pa, τ = 60 sec. 1 – (211) Nb, 2 – (200) Nb, 3 – (211) Nb$_3$Sn, 4 - (200) Nb$_3$Sn, 5 - (211) Nb$_3$Sn + (110) Nb, 6 - (320) Nb$_3$Sn, 7 - (321) Nb$_3$Sn

Films, condensed at T$_S$ = 700$^0$ C and U = 1000V, are multiphase, textured, which contain all the known compounds of Nb$_3$Sn, Nb$_6$Sn$_5$, and NbSn$_2$ (Fig.16). This gives us ground to state that at the nanoscale there is no strict implementation of the stoichiometric relations between the components in the formed multi-component system.

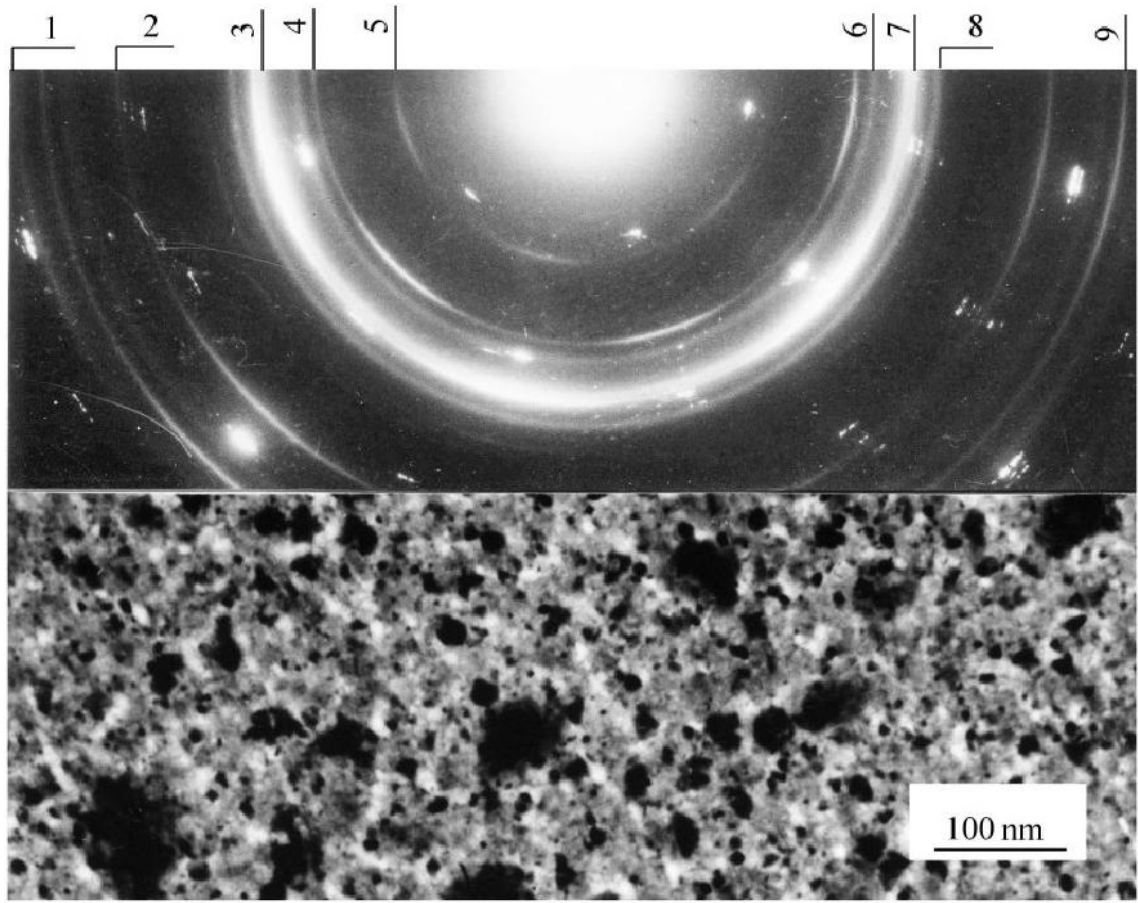

Fig.16. Phase composition of Nb-Sn alloy, formed at U = - 1000 V, p = 9 $10^{-5}$ Pa, $T_S$ = $700^0$ C.
1 – (400) $Nb_3$ Sn; 2 – (305), (244) $Nb_6$ $Sn_5$ + (624), (133),(206) Nb $Sn_2$ ; 3 - (800), (711) Nb $Sn_2$ + (026) $Nb_6$ $Sn_5$ + (210) $Nb_2$ Sn; 4 - (113) Nb $Sn_2$ + (200), (105) $Nb_6$ $Sn_5$; 5 - (022) $Nb_3$ Sn; 6 - (220), (602) Nb $Sn_2$ + (130), (033) $Nb_6$ $Sn_5$ + (200) $Nb_3$ Sn; 7 - (040), (222) (017) (204) $Nb_6$ $Sn_5$; 8 - (422), (404) Nb $Sn_2$ + (211) $Nb_3$ Sn + (141), (215) (107) $Nb_6$ $Sn_5$; 9 – (026), (040) Nb $Sn_2$ + (321) $Nb_3$ Sn

## 5. PROCESSES ON THE TARGET DURING ITS IRRADIATION BY ACCELERATED IONS OF NIOBIUM

Constructively the target is made as a two-layer one. Tin was melted on the copper water-cooled base of the target. Lateral surfaces of the target and its holder are protected by grounded screens. Magnetic field on the target surface and in close proximity had not been used. Sputtering of the target was performed at a temperature close to the room temperature. When the initial stream, emitted by the niobium cathode, contacts the grounded target (taking into account the particles, generated by vacuum arc [40]), its possible to predict how the following processes will occur: low-energy implantation of high energy atoms to a depth of several mono-layers, surface deposition of micro-droplets of neutral and ionized atoms of niobium, whose energy is insufficient for the implementation into the crystal lattice; reflection of fast particles from the surface of the atoms. Since the reflection of the plasma stream, generated by the niobium cathode, less than one, the surface of the target at its zero potential, will be covered with a niobium film quickly enough.

Increasing the accelerating potential causes a rise in the concentration of niobium atoms in the resulting stream, emitted by the target. A dynamic equilibrium takes place at some critical value of the accelerating potential of the target: the number of niobium particles incident on the target becomes equal to the number of particles, emitted by the target. This means that when using a pure ion beam, the partial sputtering coefficient of niobium reaches its constant value equal to one. Increasing the accelerating potential gives rise to the second component in the resulting stream- the atomic flux of tin. Increase of the second component determines the further increase in the density of the sputtered flux.

Experimentally established that the concentration of tin in condensing stream decreases with increasing the angle φ (Fig.12). Also the relative change in the angular dependence of the concentration is less at lower values of the accelerating potential of the target. It does not contradict with the known outcome dependence of the atomic components from the two-component target at various energies and irradiation angles of sputtering ions [41].

Lets calculate the integral sputtering coefficient at an average energy of bombarding ions $E_1$=3 keV and $E_2$=9 keV. Rate of alloy film growth and concentration of components in the film are determined from the dependences, presented on Fig.11, 12. Using the results of computer calculation of the self-sputtering coefficient of niobium ($S_0$) and applying the program SRIM-2003.10, we obtain, in accordance with the above energies of bombarding ions, that $S_{01} = 2.4$, $S_{02} = 4,1$. According to (30) we get that $K_1 = 0,34$, $K_2 = 0,27$. From (26) we obtain that integral coefficients of sputtering are: $S_1 \approx 2,8$ and $S_2 \approx 5,2$. From (19), neglecting the reflection of ions, we find that the partial coefficient of sputtering of niobium is $S_{Nb,p} \approx 1,1$ ($\gamma \approx 0,9$). While the surface concentration of niobium is: $c_{\delta 1} \approx 0.46$, $c_{\delta 2} \approx 0.28$. From (24) we find the ratio of the volume concentration of niobium in the target to the surface concentration ($\xi = c_\Delta / c_\delta$): $\xi_1 \approx 0.55$, $\xi_2 \approx 0.57$, i.e. the surface of the target is enriched with niobium.

Lets note the following features of the sputtering process when using the method of transformation of the ion flux. The target should be seen as a flat electric probe, located in plasma, during the ignition of arc discharge (as it follows from Fig.1). Therefore, the ions are accelerated within the modified layer, in which the electric field differs from zero. The initial kinetic energy of a directed motion of ions, generated by cathode of vacuum arc, is the same for all ions, irrespective of the degree of their ionization [10]. This implies that the component of ion velocity parallel to the target surface is a constant. Therefore (see Fig.1), the angle of incidence of the accelerated ions on a flat surface of the target, which is located at $45^0$ relatively to the direction of plasma flow, will decrease with increasing the target potential. Based on the above, during the computer calculation of the integral sputtering coefficient of the alloy, the angle of irradiation was varied in such a way that the partial sputtering coefficient of niobium was equal to $1/\gamma$. Using the values of the surface concentration of the components, calculated from experimental data, the density and binding energies of the atoms (surface and volume) were averaged. Moreover, when evaluating the binding energy of atoms in the alloy, based on the sublimation energy of the elementary substances, we must consider the energy released during the displacement of the components (see for example [42]). However, practical work on calculation of the sputtering coefficient, using SRIM-2003.10 program, allowed us to establish that the result of computer calculations is most sensitive to changes in surface concentration of components and angle of incidence of bombarding ions. At the same time result of computer calculation weakly depends on changes in the threshold displacement energy, as well as surface and volume binding energy. 100% increase in binding energy does not significantly affect the final result. Therefore, the averaging of the binding energy in the alloy, using the appropriate values of pure elements in the approximation of additive contribution of the components, does not lead to a significant error.

The calculated values of integral sputtering coefficient such as $S_1 \approx 4,2$, when $E_1 = 3$ keV and $S_2 \approx 7,4$, when $E_2 = 9$ keV were overestimated compared with the experimentally determined values given above.

Lower experimentally determined value of integral sputtering coefficient is apparently due to the development of the relief of the target surface. The development of the relief was identified in the study of morphology of the target surface in an electron scanning microscope (Fig.17) after the end of consecutive series of experiments on its dispersal, when accelerating potentials are $U = (3; 1; 0,7; 0,5; 0,4)$ kV.

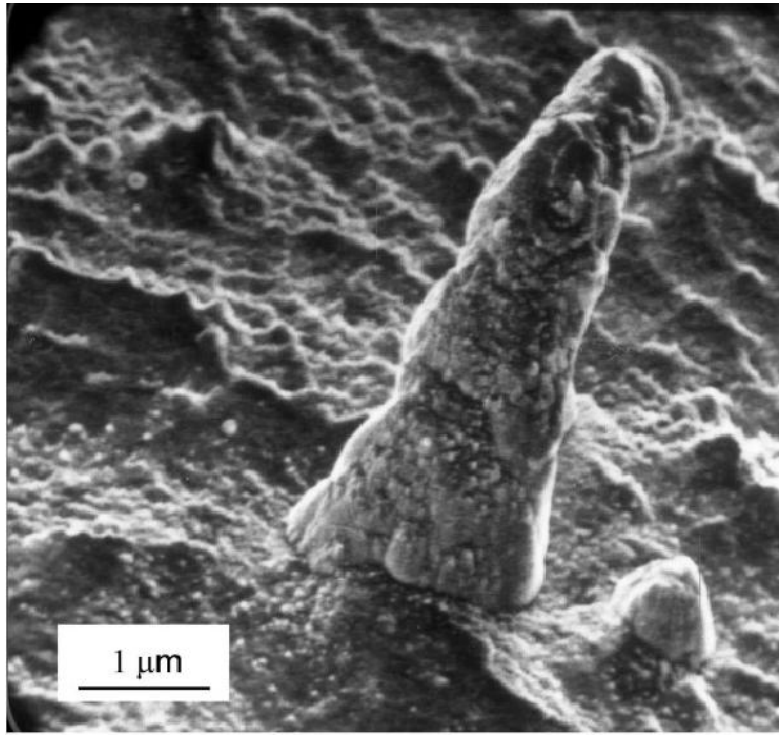

Fig.17. The development of the surface relief of tin target under accelerated Nb ion bombardment.

We recall that, in general, this work is devoted to the application of the method of transformation of ion-plasma streams for the formation of films and coatings and investigation of their structure and properties. Qualitative study of processes on the target was directed at attracting of patterns of development of surface relief (established by other authors) in order to explain the features of film growth from the sputtered streams. Lets make some assumptions about the initial phase of the development of the target surface relief, which may be associated with the following circumstances. Niobium atoms, by bombarding a tin target, form in the implantation layer (Fig.1) some local areas, that differ by stress state. First- are the areas with a high concentration of vacancies and their complexes (including the emerging dislocation subtraction loops), characterized by increased level of tensile micro-stresses. Second - areas of crystal lattice of the tin, containing implementation type defects, including the dislocation implantation loops. These areas are characterized by increased level of compressive micro-stresses (areas corresponding to the final stages of the development of collision cascade). The occurrence of radiation-induced diffusion in the layer of introduction (diffusion rate is proportional to the concentration of created point defects) may cause local segregation of niobium in the tin lattice with the formation of nuclei of new phase. This means that, from the microscopic point of view, there is some heterogeneity of the concentration of target components in the near- surface layer. This heterogeneity correlates with the level and the sign of the local stress state of the crystal lattice of the matrix. Natural to expect that the ion etching of such regions occurs at different rates. The spatial distribution of small volumes, the stresses in which differ by the sign, have some statistical variations. Due to the statistical variations there is a possibility of formation of regions, characterized by high level of compressive micro-stresses with the configuration corresponding to the stress state of all-round compression. During the sputtering process the target surface approaches to such areas, and the relaxation of stress state will proceed with the emergence of local mass transfer process, which on the target surface appears in form of whisker nucleus. Mechanism of whisker-formation under the influence of stresses was discussed, for example, in paper [43]. All of the above, first of all, concerns the mechanisms of development of surface relief on the nano-scale, which are addressed in the literature least of all. During irradiation of tin target, located within the zone of sight from the cathode of arc source, we should expect phenomena, which is described by proper and improper mechanisms of development of surface relief at the micron level [41]. This is due to the fact that part of the target surface of tin is covered with micro-droplets of niobium of micron and of sub-micron size. As it is well known [12], the sputtering coefficients of these individual substances significantly differ.

During the experiments on the sputtering of the target, the submission of an accelerating voltage was carried out in following sequence: from the larger value to the smaller. Obviously, that the development of an equilibrium relief of target surface was achieved at the accelerating potential of 3 kV. The time of preliminary bombardment of the target in order to reach the stationary sputtering regime was selected based on the dependence (10).

Stationary regime of targets sputtering corresponds to the dynamic equilibrium of the process of formation and smoothing of the micro-size roughness. Bombardment of the target with low-energy ions at high doses of radiation, on the micron level often leads to the formation of quasi-periodic wave-like structure, which is more clearly manifests for semiconductors and insulators. In this case the formed combs are located perpendicular to the direction of the beam [41]. It does not contradict the general pattern of formation of quasi-periodic structures in the stationary state during the occurrence in an open systems of non-equilibrium processes with the inverse link. Here the inverse link is manifested in the form of competing processes of sputtering and condensation of targets materials. During the sputtering process combs change their height relatively to the mean value. We can also expect the movement of the combs parallel to the targets surface during the sputtering of their lateral surfaces at different speed.

Lets estimate the effect of roughness on the occurrence of possible concentrators of intensity of electric field on the surface of the target. While the target, located in the plasma of arc discharge, will be regarded as flat Longmuir's probe. From [16, 44, 45] we can write the expressions for the ion current of saturation on the probe and width of the layer ($D$) of the volume charge of ions, which separates the target from plasma boundary.

$j_i = 8{,}9 \cdot 10^{-14} n \sqrt{E/\mu}$ (a), where $n$ - ion density in the plasma (cm$^3$), $E$ - average electron energy (eV), $\mu$ - molar mass (g/mol), $j_i$ - density of ion current (A/cm$^2$);

$D = \sqrt{4\varepsilon_0/9j_i}\,(2eN_0/\mu)^{1/4} U^{3/4}$ (b), where $N_0$-Avogadro number, $\mu$ - molar mass, $U$ - accelerating voltage on the target (expression (b) is given in SI system). Experimental value of density of the ion stream onto the target $j_i \approx 1{.}3\,10^{-2}$ A/cm$^2$, when $E = 3$ keV, therefore from (a) we obtain that $n \approx 8\,10^{11}$ cm$^{-3}$. From (b) we find that D varies from ~ 0,6mm to ~ 3mm, when potential of the target changes from 0,4 kV to 3 kV. We note that for the above data the Debye radius is $r_D$ ~ 10 μm, i.e. $r_D / D \ll 1$ ($r_D = 5 \cdot 10^2 \sqrt{E/n}$ [45], where $E$ is in eV, and $n$ is in cm$^{-3}$).

From the obtained values of thickness of the positive volume charge layer, compared with a typical dimensions of the targets micro-roughness height (see Fig.17), follows that the ions are accelerated in the electric field of the flat target. We can neglect the concentrators of electric field intensity in the initial moment of acceleration. Surface relief plays a significant role in the final stage of acceleration in the moment of collision of accelerated particle with the surface, defining the real (local) angle of bombardment. For simplicity we assume that, all the combs have the same angle at the top, then when shading does not occur, for the sputtering angle ($\psi$) of the combs surfaces, which are facing the cathode (odd numbers) and distant from it (even numbers), respectively can be written: $\psi_{2n+1} = 90 - (\theta + \alpha)$, $\psi_{2n} = 90 - (\theta - \alpha)$, where $\theta$ - half of the angle at the top of the comb, $\alpha$ - angle of bombardment of the flat target, which depends on the accelerating voltage. Since the initial kinetic energy of niobium ions is about 116 eV [10], we find that $\alpha$ for ions with an average charge $3\,|e|$ decreases from ~ $17^0$ to ~ $6^0$ when accelerating voltage increases from 0,4 kV to 3 kV. Based on actual charges of the ions $\alpha$ varies in the broader margins. Considering all the above and taking into account the fact that for the accelerated ions with an average mass the sputtering coefficient reaches maximum values in the range of angles from $63^0$ to $85^0$ [16], we can explain the differences in the angular dependence of film deposition rate (see Fig.11). Indeed, the increase of α , and consequently ψ$_{2n}$ under the decrease in the targets accelerating potential, leads to an increase in the relative value of the stream, emitted by the lateral surface with even numbers, which maximum value is attained at some angle φ (see Fig.1). Should be borne in mind that some fraction of reflected ions is present in the total beam, at the same time it is obvious that the relative contribution of the reflected particles into the resulting beam (which depends on the accelerating potential of the target) is more on the surface (2n).

Computer calculation of sputtering coefficient enables to estimate the mean value of the angle θ, under averaging of sputtering angle: $\langle \psi \rangle = \sum_{n=0}^{N} 1/N \langle \psi_{2n+1} + \psi_{2n} \rangle \approx 1/2 \langle \psi_1 + \psi_2 \rangle$.

When the energy of bombarding niobium particles is E= 9keV, varying ψ, it was found that the sputtering angle $\langle \psi \rangle \approx 36^0$, hence $\langle \theta \rangle \approx 54^0$.

### 6. FINDINGS

Processes of vacuum condensation of the films and ion sputtering of the substance are non-equilibrium processes in an open systems. Experimentally determined macroscopic physical values, described in the framework of classical physics, which characterize the system, experience the fluctuations of different levels.

Final structure and physical-chemical properties of the formed film system depend on the evolution of the system in an open state, which has a probabilistic nature. The degree of dis-equilibrium of the processes in open systems is determined by the kinetic parameters, among which the flux density of condensed particles and their energy are most important.

Experimental observation of a certain parameter of this open system in the stage of formation involves averaging of it over a certain sufficiently large ensemble of particles. Under stationary conditions of formation the level of fluctuations of the investigated parameter correlates with a certain finite size of the considered region, which extent depends on the degree of dis-equilibrium of the open system. In this sense, we can speak about the uncertainty of the physical quantity. For example, in case of transformation of the ion-plasma stream, the maximum uncertainty inherits in the stress state and the concentration of components in the layer of implantation of the sputtered target. Completion of the formation of a certain system implies the termination of mass transfer, which leads the formed system into the closed state, characterized by a significantly lower degree of an external influence.

Transition of the system into the close state changes the values of physical quantities, which describe the state of this system. The transition occurs from the dynamic, stationary values to the quasi-equilibrium, including the meta-stable values. While a certain level of fluctuations of physical quantities is frozen in the system. We can conclude that, at the nano-scale in the formed multi-component systems essentially impossible to obtain strict compliance of stoichiometric ratio of the components. The degree of remoteness of the real average concentrations from their values in fixed nano-volume depends on the method of formation of thin films.

Tendency of the system to the equilibrium with the minimization of free energy involves the occurrence of diffusion processes, which transport link under normal conditions are the vacancies. Decrease in their concentration below the thermodynamic equilibrium value fixes the system in a meta-stable state, which is typical for films, formed in the energy range $E_0 < E \leq 2E_d$, $\langle E \rangle \sim E_d$.

### 6.1. CONCLUSIONS

1. On the example of the formation of niobium films and the niobium-tin films the application of the method of transformation of the ion-plasma stream, generated by the cathode of vacuum-arc device, was demonstrated.

2. Experimentally established that under the condensation temperature of 1/3 $T_m$ in the energy range $E_0 < E \leq 2E_d$, $\langle E \rangle \sim E_d$ of deposited particles, highly stable mono-component films, containing an increased level of macro-stresses, are formed.

3. Elementary phenomenological model was proposed, that explains the physical essence of the formation of binary alloys during the sputtering of a conducting target with accelerated ions of metals.

4. The role of thermodynamic and kinetic factors, which manifests itself in their relationship, in the formation of specific structure and phase composition of films was demonstrated for mono-component and bi-component films.

5. Distinction between radiation-induced (RID) and condensation-induced (CID) diffusion was carried out: RID - $E \gg E_d$; CID (a) - $E \gtrsim E_0$, $\langle E \rangle \sim kT$ (vacancy mechanism), (b) - $E_{i,f} < E \leq 2E_d$, $\langle E \rangle \sim E_d$ (interstitial mechanism).

6. Identified long-range effect in the homo-epitaxial formation of coatings is a consequence of the condensation-induced diffusion of interstitial mechanism in the film-substrate system. Its practical application in corrosion protection of massive samples was suggested.